\documentclass[10pt,journal,compsoc]{IEEEtran}

\ifCLASSOPTIONcompsoc
  \usepackage[nocompress]{cite}
\else
  \usepackage{cite}
\fi
\ifCLASSINFOpdf
\else
\fi
\usepackage{url}
\usepackage[hidelinks]{hyperref}
\usepackage[utf8]{inputenc}
\usepackage{graphicx}
\usepackage{amsmath}
\usepackage{amsfonts}
\usepackage{booktabs}
\usepackage{graphicx}
\usepackage{algorithm}
\usepackage[noend]{algpseudocode}

\begin{document}
%
\title{GloDyNE: Global Topology Preserving Dynamic Network Embedding}
%
%
%
%

\author{Chengbin~Hou, Han~Zhang, Shan~He,
        and~Ke~Tang,~\IEEEmembership{Senior Member,~IEEE}
\IEEEcompsocitemizethanks{\IEEEcompsocthanksitem C. Hou and K. Tang are with the Guangdong Provincial Key
Laboratory of Brain-Inspired Intelligent Computation, Department of Computer Science and Engineering, Southern University of Science and Technology, Shenzhen 518055, China. (Corresponding author: Ke Tang.)\protect\\
E-mail: chengbin.hou10@foxmail.com~and~tangk3@sustech.edu.cn
\IEEEcompsocthanksitem C. Hou, H. Zhang, and S. He are with the School of Computer Science, University of Birmingham, Birmingham, B15 2TT, United Kingdom.\protect\\
E-mail: hxz325@cs.bham.ac.uk~and~s.he@cs.bham.ac.uk
}
\thanks{Manuscript accepted 16 December 2020, IEEE-TKDE.\protect\\
\scriptsize{© 2020 IEEE. Personal use of this material is permitted. Permission from IEEE must be obtained for all other uses, in any current or future media, including reprinting/republishing this material for advertising or promotional purposes, creating new collective works, for resale or redistribution to servers or lists, or reuse of any copyrighted component of this work in other works.} 
}}

%
%

\markboth{Journal of \LaTeX\ Class Files,~Vol.~XX, No.~XX, Dec~2020}%
{Shell \MakeLowercase{\textit{et al.}}: Bare Advanced Demo of IEEEtran.cls for IEEE Computer Society Journals}
%



\IEEEtitleabstractindextext{%
\begin{abstract}
Learning low-dimensional topological representation of a network in dynamic environments is attracting much attention due to the time-evolving nature of many real-world networks. The main and common objective of Dynamic Network Embedding (DNE) is to efficiently update node embeddings while preserving network topology at each time step. The idea of most existing DNE methods is to capture the topological changes at or around the most affected nodes (instead of all nodes) and accordingly update node embeddings. Unfortunately, this kind of approximation, although can improve efficiency, cannot effectively preserve the global topology of a dynamic network at each time step, due to not considering the inactive sub-networks that receive accumulated topological changes propagated via the high-order proximity. To tackle this challenge, we propose a novel node selecting strategy to diversely select the representative nodes over a network, which is coordinated with a new incremental learning paradigm of Skip-Gram based embedding approach. The extensive experiments show GloDyNE, with a small fraction of nodes being selected, can already achieve the superior or comparable performance w.r.t. the state-of-the-art DNE methods in three typical downstream tasks. Particularly, GloDyNE significantly outperforms other methods in the graph reconstruction task, which demonstrates its ability of global topology preservation.
\end{abstract}
}

\maketitle

\IEEEdisplaynontitleabstractindextext

%
\IEEEpeerreviewmaketitle

\ifCLASSOPTIONcompsoc
\IEEEraisesectionheading{\section{Introduction}\label{sec:introduction}}
\else
\section{Introduction} \label{Sec1}
\fi
\IEEEPARstart{T}{he} interactions or connectivities between entities of a real-world complex system can be naturally represented as a network (or graph), e.g., social networks, biological networks, and sensor networks. Learning topological representation of a network, especially low-dimensional node embeddings which encode network topology therein so as to facilitate downstream tasks, has received a great success in the past few years \cite{cui2018survey,hamilton2017representation,goyal2018graph}.

Most previous Network Embedding methods such as \cite{perozzi2014deepwalk,tang2015line,cao2015grarep,grover2016node2vec,ou2016asymmetric} are designed for \textit{static networks}. However, many real-world networks are dynamic by nature, i.e., edges might be added or deleted between seen and/or unseen nodes as time goes on. For instance, in a wireless sensor network, devices will regularly connect to or accidentally disconnect from routers; in a social network, new friendships will establish between new users and/or existing users. Due to the \textit{time-evolving nature} of many real-world networks, Dynamic Network Embedding (DNE) is now attracting much attention \cite{zhu2016scalable,li2017attributed,goyal2017dyngem,zhu2018high,zhang2018timers,du2018dynamic,zhou2018dynamic,chen2018scalable,mahdavi2018dynnode2vec,singer2019node,trivedi2019dyrep}. The main and common objective of DNE is to efficiently update node embeddings while preserving network topology at each time step. Most existing DNE methods try to compromise between effectiveness (evaluated by downstream tasks) and efficiency (while obtaining node embeddings). The idea is to capture the topological changes at or around the most affected nodes (instead of all nodes), and promptly update node embeddings based on an efficient incremental learning paradigm.

\begin{figure*}[htbp]
    \centering
    \includegraphics[width=0.96\textwidth]{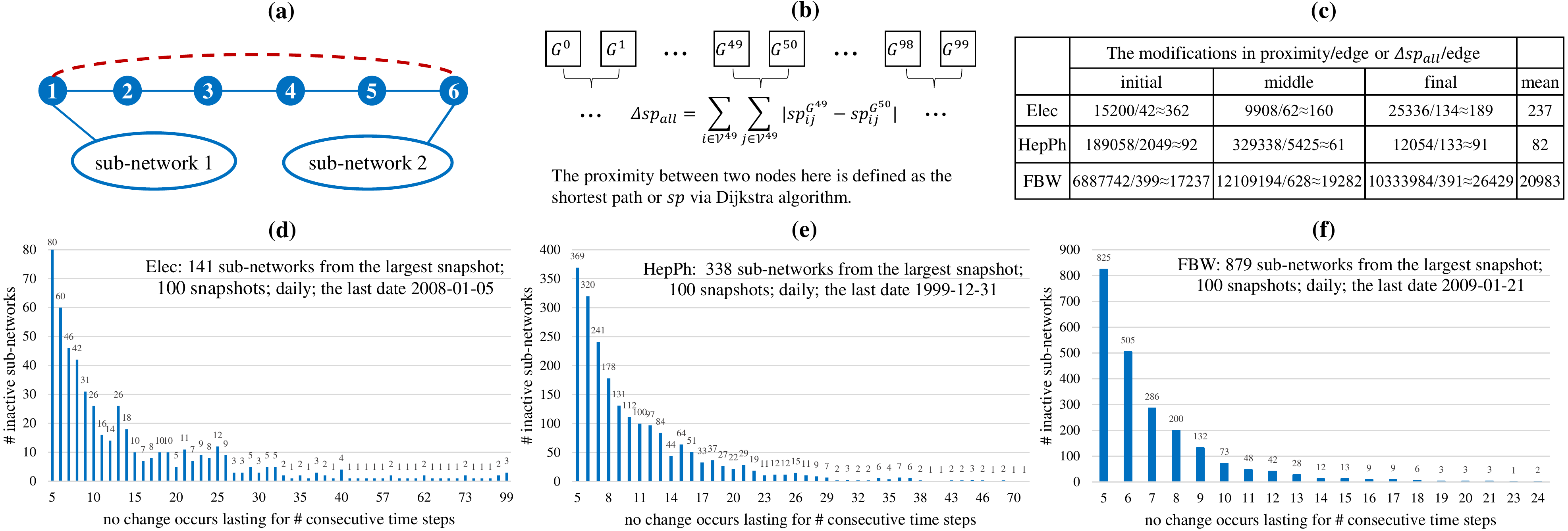}
    \caption{
    a) A change (new edge in red) affects all nodes in the connected network via high-order proximity. The proximity of nodes 1-6 becomes $1^{st}$ order from $5^{th}$ order, nodes 2-6 becomes $2^{nd}$ order from $4^{th}$ order, etc. The proximity of any node in sub-network 1 to any node in sub-network 2 is reduced by 5 orders. 
    b-c) How to calculate the modifications of the proximity between two snapshots, and the results show the modifications caused by a single edge can be very large in the real-world dynamic networks.
    d-f) The real-world dynamic networks have some inactive sub-networks, e.g., defined as no change occurs lasting for at least 5 time steps. The sub-networks, each of which has about 50 nodes, are obtained by applying METIS algorithm \protect\cite{karypis1998fast} on the largest snapshot of a dynamic network.
    The details of the three dynamic networks are described in Section \ref{datasets}.}
    \label{Fig1}
\end{figure*}

Unfortunately, this kind of approximation, although can improve the efficiency, cannot effectively preserve the global topology of a dynamic network at each time step. Specifically, any changes, i.e., edges being added or deleted, would affect all nodes in a connected network and greatly modify the proximity between nodes over a network via the high-order proximity as illustrated in Figure \ref{Fig1} a-c). On the other hand, as observed in Figure \ref{Fig1} d-f), the real-world dynamic networks usually have some \textit{inactive sub-networks} where no change occurs lasting for several time steps. Putting both together, the existing DNE methods that focus on the most affected nodes (belonging to the active sub-networks) but do not consider the inactive sub-networks, would overlook the accumulated topological changes propagating to the inactive sub-networks via the high-order proximity. 

To tackle this challenge, the proposed DNE method--\underline{Glo}bal topology preserving \underline{Dy}namic \underline{N}etwork \underline{E}mbedding (GloDyNE) first partitions a current network into smaller sub-networks where one representative node in each sub-network is selected, so as to ensure the \textit{diversity} of selected nodes. The representative node for each sub-network is sampled via a probability distribution over all nodes within each sub-network, such that a higher probability is assigned to a node with the larger accumulated topological changes. After that, GloDyNE captures the latest topologies around the selected nodes by truncated random walks \cite{perozzi2014deepwalk}, and then promptly updates node embeddings based on the Skip-Gram Negative Sampling (SGNS) model \cite{mikolov2013distributed} and an incremental learning paradigm. 

The contributions of this work are as follows.
1) We demonstrate the existence of inactive sub-networks in real-world dynamic networks. Together with the propagation of topological changes via the high-order proximity, we find the issue of global topology preservation for many existing DNE methods.
2) To better preserve the global topology, \textit{unlike all previous DNE methods}, we propose to also consider the accumulated topological changes in the inactive sub-networks. A novel node selecting strategy is thus proposed to diversely select the representative nodes over a network. 3) We further develop a new DNE method or framework, namely GloDyNE, which extends the random walk and Skip-Gram based network embedding approach to an incremental learning paradigm with a free hyper-parameter for controlling the number of selected nodes at each time step. 4) The extensive empirical studies show the superiority of GloDyNE compared with the state-of-the-art DNE methods in terms of both effectiveness and efficiency, as well as verify the usefulness of some special designs of GloDyNE, such as the node selecting strategy and the free hyper-parameter.

The remainder of the paper is organized as follows. We first review the related works in Section \ref{Sec2}, and then formally give the definition of DNE problem in Section \ref{Sec3}. In Section \ref{Sec4}, we present GloDyNE step by step, as well as its pseudocode and theoretical time complexity. The empirical studies are reported and discussed in Section \ref{Sec5}. In particular, Section \ref{Sec5.2} aims to compare GloDyNE with other DNE methods, whereas Section \ref{Sec5.3} tries to investigate GloDyNE itself. And finally, we conclude this work in Section \ref{Sec6}.

\section{Related Works} \label{Sec2}
\subsection{Static Network Embedding} To learn low-dimensional topological representation of a network in dynamic environments, one naive solution is to treat the snapshot of a dynamic network at each time step as a static network, so that a static Network Embedding method such as \cite{perozzi2014deepwalk,tang2015line,grover2016node2vec,ou2016asymmetric} can be directly applied to learn node embeddings for each snapshot. As reported in recent DNE works \cite{zhang2018timers,zhu2018high,du2018dynamic,zhu2016scalable}, this naive solution obtains superior results compared with some DNE methods. One possible reason is that this solution does not suffer the aforementioned issue of global topology preservation. However, it is time-consuming \cite{du2018dynamic,zhu2018high}, and thus may not satisfy the requirement of promptly updating embeddings for some downstream tasks \cite{chen2018scalable,yu2018netwalk}.

\subsection{Dynamic Network Embedding} To compromise between effectiveness and efficiency, most existing DNE methods try to capture the topological changes at or around the most affected nodes (instead of all nodes or edges), and promptly update node embeddings based on an incremental learning paradigm. BCGD \cite{zhu2016scalable} aims to minimize the loss of reconstructing the network proximity matrix using the node embedding matrix with a temporal regularization term, and it is optimized by the Block-Coordinate Gradient Descent algorithm. Particularly, this work further offers an efficient solution--BCGD-incremental that only updates the most affected nodes' embeddings based on their previous embeddings. DynGEM \cite{goyal2017dyngem} and NetWalk \cite{yu2018netwalk} both utilize an auto-encoder with some regularization terms for modeling. They continuously train the model inherited from the last time step, so that the model converges in a few iterations thanks to the knowledge transfer from previous models. 
To efficiently cope with dynamic changes at each time step, some DNE methods \cite{du2018dynamic,mahdavi2018dynnode2vec} propose an incremental version of the Skip-Gram model \cite{mikolov2013distributed} to update embeddings based on the most affected nodes. Likewise, DHEP \cite{zhu2018high} extends HOPE \cite{ou2016asymmetric} to an incremental version by modifying the most affected eigenvectors using the matrix perturbation theory.

Apart from above DNE methods with the trade-off between effectiveness and efficiency, some DNE methods \cite{zhou2018dynamic,singer2019node,trivedi2019dyrep} aim to further improve effectiveness without considering the efficiency. For example, tNE \cite{singer2019node} runs a static Network Embedding method to get node embeddings at each time step, and then employs Recurrent Neural Networks among them (for better exploiting the temporal dependence and hence may further improve effectiveness) to obtain the final node embeddings at each time step. 

Unlike the previous DNE works that mainly consider the most affected nodes or sub-networks, this work proposes to also consider the accumulated topological changes in inactive sub-networks to better preserve the global topology of a dynamic network. Moreover, this work focuses on the network topology, but does not include node attributes \cite{yang2015network,gao2018deep,huang2017accelerated}, which will be left as the future work.

\section{Notation and Problem Definition} \label{Sec3}
\textit{Definition 1. A Static Network.} Let $G=(\mathcal{V},\mathcal{E})$ be a static network where $\mathcal{V}$ denotes a set of $|\mathcal{V}|$ nodes or vertices, and $\mathcal{E}$ denotes a set of $|\mathcal{E}|$ edges or links. The adjacency matrix of $G$ is denoted as $W\in \mathbb{R}^{|\mathcal{V}|\times |\mathcal{V}|}$ where $w_{ij}$ is the weight of edge $e_{ij}$ between a pair of nodes $(v_i,v_j)$, and if $w_{ij}=0$, there is no edge between the two nodes. 

\textit{Definition 2. A Dynamic Network.} A dynamic network $\mathcal{G}$ is represented by a sequence of snapshots $G^t$ taken at each time step $t$, i.e.,  $\mathcal{G}=(G^0,G^1,...,G^t,G^{t+1},...)$. Each snapshot $G^t$ can be treated as a static network.

\textit{Definition 3. Static Network Embedding.} The static network embedding aims to find a mapping function $Z=f(G)$, where $Z\in\mathbb{R}^{|\mathcal{V}|\times d}$, $d \ll |\mathcal{V}|$, and each row vector $Z_i\in\mathbb{R}^d$ is the node embedding for $v_i$, such that the pairwise similarity of node embeddings in $Z$ best preserves the pairwise topological similarity of nodes in $G$. 

\textit{Definition 4. Dynamic Network Embedding.} The DNE problem, under an incremental learning paradigm, can be defined as $Z^t=f^t(G^t,G^{t-1},f^{t-1},Z^{t-1})$ where $Z^t\in \mathbb{R}^{|\mathcal{V}^t|\times d}$ is the latest node embeddings, $f^{t-1}$ and $Z^{t-1}$ are the model and embeddings from the last time step respectively. The main objective of DNE in this work is to efficiently update node embeddings at each current time step $t$, such that the pairwise similarity of node embeddings in $Z^t$ best preserves the pairwise topological similarity of nodes in $G^t$. 

\textit{Definition 5. Sub-networks of A Snapshot.} Let $G^t_k$ denote the $k^{th}$ sub-network in a snapshot $G^t$. All sub-networks of a snapshot $G^t$, after network partition \cite{bulucc2016recent}, should be non-overlapping, i.e, $\mathcal{V}^t_{m} \cap \mathcal{V}^t_{n} = \emptyset$, $\forall~~m \neq n$. And their node sets should satisfy $\mathcal{V}^t = \bigcup\nolimits_{k} \mathcal{V}^t_k$.

\section{The Proposed Method} \label{Sec4}
The proposed DNE method--GloDyNE consists of four essential components. We will introduce them step by step in Section \ref{Sec4.1}. Intuitively, Step 1 and 2 ensure the selected nodes \textit{diversely distributed over a network}, and meanwhile, bias to the nodes with the larger accumulated topological changes in each sub-network. Step 3 encodes the latest topologies around the selected nodes into random walks (i.e., node sequences), which are then decoded by a sliding window and the SGNS model for incrementally training node embeddings as described in Step 4. Note that, these four steps are repeatedly executed at each time step.

\subsection{Method Description} \label{Sec4.1}
\subsubsection{Step 1. Partition A Network} \label{Step1}
In order to realize inactivate sub-networks of a snapshot $G^t$, it is needed to divide $G^t$ into sub-networks $G^t_1, G^t_2, ..., G^t_K$ where $K$ is the number of sub-networks of a snapshot. The sub-networks are desirable to be non-overlapping and to cover all nodes in the original snapshot as defined in Definition 5, so that the later Step 2 can select unique nodes from each sub-network and the later Step 3 is easier to explore the whole snapshot $G^t$ based on the selected nodes from each sub-network. A network partition algorithm \cite{bulucc2016recent} is therefore used to achieve the desirable goals. The most common objective function is to minimize the edge cut i.e.
\begin{equation}
  \min~~\sum\limits_{1 \le m,n \le K} {\{ w_{i,j}^t| v_i^t \in \mathcal{V}_m^t,v_j^t \in \mathcal{V}_n^t, (v_i^t,v_j^t) \in \mathcal{E}^t \}}
\label{eq1}
\end{equation}

\noindent where the subscripts $i,j$ indicate node ID and $m,n$ indicate sub-network ID. Note that Eq. (\ref{eq1}) should subject to two constraints $\mathcal{V}^t_{m} \cap \mathcal{V}^t_{n} = \emptyset$, $\forall~~m \neq n$, and $\mathcal{V}^t = \bigcup\nolimits_{k} \mathcal{V}^t_k$ for the reasons as discussed above. 

Moreover, an additional constraint of the balanced sub-networks is introduced to let the number of nodes be similar among all sub-networks, so as to facilitate the later steps to fairly explore all sub-networks and hence better preserve the global topology. The third constraint about the balanced sub-networks can be defined as
\begin{equation}
\forall k  \in \{ 1,...,K\} ,|\mathcal{V}^t_k| \le (1 + \epsilon) \frac{|\mathcal{V}^t|}{K}
\label{eq2}
\end{equation}

\noindent where $|\mathcal{V}^t_k|$ is the number of nodes in $G_k^t$ and $\epsilon$ is the tolerance parameter. Note that, if $\epsilon$ is 0, network partitions are perfectly balanced. In practice, $\epsilon$ is set to a small number to allow a slight violation. However, such a $(K,\epsilon)$ balanced network partition is an NP-hard problem \cite{bulucc2016recent}. In order to address this problem, METIS algorithm \cite{karypis1998fast} is employed. There are roughly three steps. First, the coarsening phase, the original network is recursively transformed into a series of smaller and smaller abstract networks, via collapsing nodes with common neighbors into one collapsed node until the abstract network is small enough. Second, the partition phase, a $K$-way partition algorithm is applied on the smallest abstract network to get the initial partition of $K$ sub-networks. Third, the uncoarsening phase, it recursively expands the smallest abstract network back to the original network, and meanwhile recursively swaps the collapsed nodes (or the original nodes lastly) at the boarder of sub-networks between two neighboring sub-networks, so as to minimize the edge cut as describe in Eq. (\ref{eq1}).

\subsubsection{Step 2. Select Representative Nodes} \label{Step2}
In order to ensure the selected nodes diversely distributed over a snapshot $G^t$, one natural idea is to select one representative node from each sub-network. As a result, the total number of selected nodes is $K$. We let $K=\alpha|\mathcal{V}^t|$, so that $\alpha$ can freely control the total number of selected nodes for the trade-off between effectiveness and efficiency.

The problem now becomes as how to select one representative node from a sub-network. According to the latest DNE works such as \cite{du2018dynamic,mahdavi2018dynnode2vec,yu2018netwalk}, the nodes affected greatly by edge streams are selected for updating their embeddings, since their topologies are altered greatly. Similarly, in this work, the representative node to be selected is biased to the node with larger topological changes. Motivating by the concept of inertia\footnote{Here the node degree is regarded as the inertia of this node.} from Physics, an efficient scoring function is designed to evaluate the accumulated topological changes of a node $v^t_i$ in a current snapshot $G^t$ as follows

\begin{small}
\begin{flalign}
\label{eq3}
S(v^t_i)
     &= \frac{|\Delta\mathcal{E}^{t}_{i}|+\mathcal{R}^{t-1}_{i}}{Deg(v^{t-1}_i)}\\\nonumber
     &= \frac{{|~\mathcal{N}({v_i^t}) \cup \mathcal{N}({v_i^{t - 1}}) - \mathcal{N}({v_i^t}) \cap \mathcal{N}({v_i^{t - 1}})}~| 
     + \mathcal{R}_i^{t-1}}{Deg(v^{t-1}_i)} &&
\end{flalign}
\end{small}

\noindent where the reservoir $\mathcal{R}^{t-1}_{i}$ stores the accumulated changes\footnote{The accumulated changes in reservoir are used to handle the case when a node has small changes at each time step for a long time, which greatly affects network topology but maybe ignored if not recorded.} of $v_i$ up to $t-1$. For simplicity, we treat $G^t$ as an undirected and unweighted network\footnote{If one wants to consider edge's weight in Eq. (\ref{eq3}), let $|\Delta \mathcal{E}_i^t|$ $=$ $\sum\nolimits_{v_j^t \in {\cal N}(v_i^t)} {|w_{i,j}^t-w_{i,j}^{t-1}|}  + \sum\nolimits_{ v_j^{t-1} \in ({\cal N}(v_i^{t - 1}) - {\cal N}(v_i^t))} {|w_{i,j}^{t-1}|}$ where the first term gives the total weight changes of $i$' neighbors presented at $t$; whereas the second term gives the total weight changes of $i$' neighbors presented at $t-1$ but not presented at $t$. The operator $|\cdot|$ on a set gives its cardinal number, and on a scalar gives its absolute value.}, so that the current changes of $v_i$ at $t$, denoted as $|\Delta\mathcal{E}^{t}_{i}|$, can be easily obtained by the set operations on neighbors of $v_i$ as shown in Eq. (\ref{eq3}), which is equivalent to count the number of the edges with node $v_i$ in the current edge streams $\Delta\mathcal{E}^{t}$. The representative node of a sub-network $G^t_k$ is then selected based on the probability distribution over its node set $\mathcal{V}^t_k$, i.e., 
\begin{equation}
    {P(v{^t_i})} = \frac{{{e^{{S(v_i^t)}}}}}{{\sum\nolimits_{v_j^t \in \mathcal{V}^t_k} {{e^{{S(v_j^t)}}}}}}~~~\forall~v^t_i \in \mathcal{V}^t_k
    \label{eq4}
\end{equation}

\noindent where $e$ is Euler's number and $S(v^t_i)$ is the score of the accumulated topological changes of node $v^t_i$ given by Eq. (\ref{eq3}). Note that, if $S(v^t_i)=0$, $e^0=1$ and $P(v^t_i) \neq 0$, so that even for an inactivate sub-network with no change in all nodes, the probability distribution over this sub-network is still a valid uniform distribution. Intuitively, within a sub-network, the higher score of a node given by Eq. (\ref{eq3}) is, the higher probability of this node will be selected as the representative node for this sub-network. Because one representative node from each sub-network is selected, all the selected nodes are therefore diversely distributed over the whole snapshot, and meanwhile, biased to the larger accumulated topological changes for each sub-network.

\subsubsection{Step 3. Capture Topological Changes} \label{Step3}
Given the selected representative nodes from Step 2, this step will explain how to capture the topological changes based on the selected nodes. As the topological changes at the selected nodes can propagate to other nodes via the high-order proximity, the truncated random walk sampling \cite{perozzi2014deepwalk} (instead of edge sampling \cite{tang2015line}) strategy is employed to capture the topological changes around (instead of at) the selected nodes. Concretely, for each selected node, $r$ truncated random walks with length $l$ are conducted starting from the selected node. For a random walk, the next node $v^t_j$ is sampled based on the probability distribution over its previous node's neighbors $\mathcal{N}(v^t_i)$, i.e.,
\begin{equation}
{P(v_j^t|v_i^t) } = \left\{ {\begin{array}{*{20}{c}}
{\frac{w_{ij}^t}{{\sum\nolimits_{v_{j'} \in \mathcal{N}(v^t_i)} w_{ij'}^t }}} &\text{ if } v_{j} \in \mathcal{N}(v^t_i)\\
0 & \text{otherwise}
\end{array}} \right.
\label{eq5}
\end{equation}

\subsubsection{Step 4. Update Node Embeddings} \label{Step4}
After Step 3, the latest topological information around the selected nodes is encoded in random walks. Step 4 aims to utilize the random walks to update node embeddings. Following \cite{perozzi2014deepwalk} and \cite{grover2016node2vec}, a sliding window with length $s+1+s$ is used to slide along each walk (i.e., node sequence), and the positive node-pair samples in a set $\mathcal{D}^t$ are built via $(v_{center+i}^t, v_{center}^t)$ where $i \in [-s,+s], i \neq 0$. As a result, the node-pair samples can encode $1^{st}\sim s^{th}$-order proximity of a given center node with another node. Note that, several network embedding works have shown the advantage of using the high-order proximity \cite{cao2015grarep,ou2016asymmetric,zhang2018arbitrary}. 

Assuming the observations of node pairs in $\mathcal{D}^t$ are mutually independent \cite{grover2016node2vec}, the objective function to maximize the node co-occurrence log probability over all node pairs in $\mathcal{D}^t$ can be written as
\begin{equation}
    \max\sum\limits_{(v^t_i,v^t_c) \in \mathcal{D}^t} {\log P(v^t_i|v^t_c)} 
    \label{eq6}
\end{equation}

\noindent where $v^t_c$ is the center node, and $v^t_i$ is another node with $1^{st}\sim s^{th}$ order proximity to $v_t^c$. Unlike \cite{perozzi2014deepwalk} which defines $P(v^t_i|v^t_c)$ as a softmax, we follow \cite{levy2014neural} to treat it as a binary classification problem, so as to further reduce the complexity. Concretely, it aims to distinguish a positive sample $(v_i^t,v_j^t) \in \mathcal{D}^t$ from $q$ negative samples $(v_i^t,v_{j'}^t)$s. The probability of observing a positive sample $(v_i^t,v_j^t)$ is
\begin{footnotesize}
\begin{equation}
P(B = 1|v_i^t,v_j^t) = \sigma {(Z_i^t\cdot Z_j^t)}{\rm{=}}\frac{{\rm{1}}}{{{1 + e^{-Z_i^t \cdot Z_j^t}}}}
\label{eq7}
\end{equation}
\end{footnotesize}

\noindent where $Z^t_i$ is the node embedding vector parameterized by the mapping function $f^t(v_i^t)$, the operator $\cdot$ represents the dot product between two vectors, and $P(B=1|v_i^t,v_j^t)$ gives the probability of a positive prediction given a positive sample $(v_i^t,v_j^t)$. Likewise, the probability of observing a negative sample $(v_i^t,v_{j'}^t)$ is
\begin{footnotesize}
\begin{equation}
P(B = 0|v_i^t,v_{j'}^t) =1- \sigma {(Z_i^t \cdot Z_{j'}^t)}{\rm{=}}\frac{{\rm{1}}}{{{1 + e^{Z_i^t \cdot Z_{j'}^t}}}} = \sigma (-Z_i^t \cdot Z_{j'}^t)
\label{eq8}
\end{equation}
\end{footnotesize}

\noindent where $P(B=0|v_i^t,v_{j'}^t)$ gives the probability of a negative prediction given a negative sample $(v_i^t,v_{j'}^t)$. The above Skip-Gram Negative Sampling (SGNS) model \cite{levy2014neural} then tries to maximize $P(B=1|v_i^t, v_j^t)$ for each positive sample in $\mathcal{D}^t$ and $P(B=0|v_i^t, v_{j'}^t)$ for the $q$ negative samples corresponding to each positive sample, i.e.,
\begin{equation}
    \max~~log\sigma (Z^t_i\cdot Z^t_j) + \sum\nolimits_{q} \mathbf{E}_{v^t_{j'}\sim P_{\mathcal{D}^t}}[log\sigma (-Z^t_i\cdot Z^t_{j'})]
\label{eq9}
\end{equation}

\noindent where $q$ negative samples are drawn from a unigram distribution $P_{\mathcal{D}^t}$ \cite{levy2014neural}. The overall objective of SGNS is to sum over all positive samples and their negative samples, i.e.,
\begin{equation}
    \max \sum_{(v_i^t,v_j^t) \in \mathcal{D}^t} \#(v_i^t,v_j^t)~~ Eq. (\ref{eq9})
\label{eq10}
\end{equation}
\noindent where $\#(v_i^t,v_j^t)$ denotes the number of times a positive sample occurs in $\mathcal{D}^t$. Intuitively, the more frequently a pair of nodes co-occurs, the closer their embeddings should be.

\begin{algorithm*}[htbp]
\caption{GloDyNE}
\label{alg1}
\textbf{Input}: snapshots of a dynamic network $G^0...G^{t-1},G^t$...; coefficient to determine the number of selected nodes $\alpha$; walks per node $r$; walk length $l$; sliding window size $s$; negative samples per positive sample $q$; embedding dimensionality $d$ \\ 
\textbf{Output}: embedding matrix $Z^t \in \mathbb{R}^{|\mathcal{V}^t| \times d}$ at each time step
\begin{algorithmic}[1] 
\For{$t=0$} 
\State conduct random walks with length $l$ starting from each node in $\mathcal{V}^0_{all}$ for $r$ times by Eq. (\ref{eq5})
\State build positive node-pair samples $\mathcal{D}^0$ based on each sliding window with size $s$ along each walk
\State initialize SGNS model $f^0_{rand}$, and train it using $\mathcal{D}^0$ with $q$ negative samples per positive sample by Eq. (\ref{eq9})
\State return $f^0$ and $Z^0$
\EndFor

\For{$t\geq1$}
\State calculate $K = \alpha |\mathcal{V}^t|$, and initialize $\Delta \mathcal{E}^t=\emptyset$ and $\mathcal{V}_{sel}^t=\emptyset$
\State partition $G^t$ into $K$ sub-networks $G_1^t,G_2^t,...,G_K^t$ by METIS based on Eq. (\ref{eq1}) and Eq. (\ref{eq2})
\State read edge streams $\Delta \mathcal{E}^t$ (or obtain it by differences between $G^{t-1}$ and $G^{t}$ if not given)
\State update reservoir dictionary via $\mathcal{R}^t_{v_i}=|\Delta\mathcal{E}^{t}_{i}|+\mathcal{R}^{t-1}_{i}$ for accumulating new changes of $v_i^t$ by Eq. (\ref{eq3})
\For{$k \in (1,...,K)$}
\State calculate a probability distribution over all nodes in a sub-network $G_k^t$ by Eq. (\ref{eq4})
\State select one representative node based on the probability distribution, and add it to $\mathcal{V}_{sel}^t$
\EndFor
\State remove selected nodes $\mathcal{V}_{sel}^t$ from the reservoir $\mathcal{R}^t$ if exists
\State conduct random walks with length $l$ starting from each node in $\mathcal{V}^t_{sel}$ for $r$ times by Eq. (\ref{eq5})
\State build node-pair positive samples $\mathcal{D}^t$ based on each sliding window with size $s$ along each walk
\State initialize SGNS model $f^{t}=f^{t-1}$, and train it using $\mathcal{D}^t$ with $q$ negative samples per positive sample by Eq. (\ref{eq9})
\State return $f^t$ and $Z^t$
\EndFor
\end{algorithmic}
\end{algorithm*}

Finally, we extend the SGNS model as described above to an \textit{incremental learning paradigm}. The overall framework of GloDyNE can be formalized as

\begin{equation}
    Z^t=\left\{
    \begin{array}{lcl}
    f^0(G^0,f^0_{rand},Z^0_{rand}) &{t=0}\\
    f^t(G^t,G^{t-1},f^{t-1},Z^{t-1}) &{t\geq1}
    \end{array} \right.
    \label{eq11}
\end{equation}

\noindent  where $f^{t-1}$ is the trained SGNS model from last time step, $Z^{t}$ is the current embedding matrix directly taken from newly trained $f^{t}$ via an index operator, and $G^{t-1}$ and $G^{t}$ are the two consecutive snapshots for generating the edge streams $\Delta \mathcal{E}^t$ if not directly given. The implementation details are presented in Section \ref{Sec4.2} and \ref{Sec4.3}.

\subsection{Algorithm} \label{Sec4.2}
The pseudocode of GloDyNE is summarized in Algorithm \ref{alg1}, and the open source code is provided at \url{https://github.com/houchengbin/GloDyNE}. 

\subsection{Complexity Analysis} \label{Sec4.3}
According to Eq. (\ref{eq11}) and Algorithm \ref{alg1}, GloDyNE consists of two stages. During the \textit{offline stage}, i.e., $t=0$, Step 3 (specifically $\mathcal{V}^t_{sel}=\mathcal{V}^0_{all}$) and Step 4 are employed to obtain the initial SGNS model and node embeddings. Lines 2-5 are indeed a static network embedding method--a modified version of DeepWalk \cite{perozzi2014deepwalk}, which trains a SGNS model instead of Skip-Gram Hierarchical Softmax (SGHS) model. Therefore, the time complexity of lines 2-5 is further reduced to $O(rlw(1+q)d|\mathcal{V}^0_{all}|)$ \cite{grover2016node2vec} where $(1+q)$ is due to one positive sample corresponding to $q$ negative samples. 

During the \textit{online stage}, i.e., $t\geq1$, steps 1-4 are employed to incrementally update the SGNS model and node embeddings. For lines 7-8 corresponding to Step 1, the time complexity is  $O(|\mathcal{V}^t|+|\mathcal{E}^t|+K log K)$ \cite{karypis1998fast} where $K=\alpha|\mathcal{V}^t|, \alpha \in(0,1)$. For lines 9-14 corresponding to Step 2, the time complexity of lines 9-10 is $O(|\Delta \mathcal{E}^t|)$; the time complexity of lines 11-13 using alias sampling method \cite{grover2016node2vec} requires $O(|\mathcal{V}^t|)$; the time complexity of line 14 is $O(\alpha|\mathcal{V}^t|)$ due to $|\mathcal{V}^t_{sel}|=K=\alpha|\mathcal{V}^t|$. For lines 14-18 corresponding to Step 3 and Step 4, similarly to lines 2-5 above, the complexity of lines 14-18 is $O(rlw(1+q)d|\mathcal{V}^t_{sel}|)$ where $|\mathcal{V}^t_{sel}|=\alpha|\mathcal{V}^t|$. Because most real-world networks are sparse, edges in a snapshot $|\mathcal{E}^t|=b_1|\mathcal{V}^t|$ such that the average degree $b_1$ is a very small number compared with $|\mathcal{V}^t|$. Besides, since the edge streams between two consecutive snapshots are often much less than the edges in the snapshot, edge streams $|\Delta\mathcal{E}^t|= b_2 |\mathcal{V}^t|$ such that $b_2<b_1 \ll |\mathcal{V}^t|$. Regarding those real-world assumptions, the overall complexity of online stage (including all four steps) at each time step can be approximated as $O(|\mathcal{V}^t|+\alpha|\mathcal{V}^t| log~ \alpha|\mathcal{V}^t|+rlw(1+q)d\alpha|\mathcal{V}^t|)$ where $\alpha \in (0,1)$ is used to control the number of selected nodes, $|\mathcal{V}^t|$ denotes the number of nodes at $t$, and others are negligible constants compared to $|\mathcal{V}^t|$. GloDyNE is thus scalable w.r.t. $|\mathcal{V}^t|$, as there is no quadratic or higher term.

\section{Empirical Studies} \label{Sec5}
\subsection{Experimental Settings}
\subsubsection{Datasets} \label{datasets}
In this work, six datasets are employed to evaluate the proposed method.
\textit{To construct the dynamic networks}, except AS733 (given as the snapshot representation), all other ones (given as the edge streams $\{(v_i, v_j, timestamp),...\}$) are constructed as follows: 1) the initial snapshot $G^0$ is built by appending all edges no later than the initial cut-off timestamp; 2) the next snapshot $G^1$ is built by appending the edges newly appeared until the next cut-off timestamp to $G^0$; 3) repeat step 2 so as to generate $G^2$, $G^3$, and so on. For each snapshot, we take out the largest connected component and treat it as an undirected and unweighted graph. According to real-world practice, the cut-off timestamp is based on the last second of a calendar day, and the gap between snapshots on a same dataset is identical. The gap for different datasets would be different due to the nature of different datasets, e.g., the gap is set to more calendar days if a dataset evolves more slowly over time.

AS733 contains 733 daily instances of the autonomous system of routers exchanging traffic flows with neighbors. Since AS733 is directly given as the snapshot representation, we directly take out the recent 21 snapshots (12/Dec./1991--01/Jan./2000) to form its dynamic network. The initial snapshot has 1476 nodes and 3123 edges, and the final snapshot has 3570 nodes and 7033 edges. The original dataset comes from \url{https://snap.stanford.edu/data/as-733.html}.

Elec is the network of English Wikipedia users vote for and against each other in admin elections. The gap between the timestamps for taking snapshots is one calendar day. We take out the recent 21 snapshots (16/Dec./2007--05/Jan./2008) to form its dynamic network. The initial snapshot has 6972 nodes and 99006 edges, and the final snapshot has 7066 nodes and 100655 edges. The original dataset comes from \url{http://konect.cc/networks/elec}.

FBW is a social network of Facebook Wall posts where nodes are the users and edges are built based on the interactions in their wall posts. The gap between the timestamps for taking snapshots is one calendar day. We take out the recent 21 snapshots (01/Jan./2009--21/Jan./2009) to form its dynamic network. The initial snapshot has 41730 nodes and 169918 edges, and the final snapshot has 43952 nodes and 182365 edges. The original dataset comes from \url{http://konect.cc/networks/facebook-wosn-wall}.

HepPh is a co-author network extracted from the papers of High Energy Physics Phenomenology in arXiv. The gap between the timestamps for taking snapshots is one month. We take out the recent 21 snapshots (Apr./1998--Dec./1999) to form its dynamic network. The initial snapshot has 11156 nodes and 611311 edges, and the final snapshot has 16913 nodes and 1194408 edges. The original dataset comes from \url{http://konect.cc/networks/ca-cit-HepPh}.

Cora is a citation network where each node represents a paper, and an edge between two nodes represents a citation. Each paper is assigned with a label (from 10 different labels) based on its field of the publication. Following \cite{singer2019node}, the gap between the timestamps for taking snapshots is one year. The 11 snapshots (1989--1999) are taken out to form its dynamic network. The initial snapshot has 348 nodes and 481 edges, and the final snapshot has 12022 nodes and 45421 edges. The original dataset comes from \url{https://people.cs.umass.edu/~mccallum/data.html}.

DBLP is a co-author network in computer science field. Each author is associated with a label (from 15 different labels). The label of an author is defined by the fields in which the author has the most publications. Following \cite{singer2019node}, the gap between the timestamps for taking snapshots is one year. The 11 snapshots (1985--1995) are taken out to form its dynamic network. The initial snapshot has 1679 nodes and 3445 edges, and the final snapshot has 25826 nodes and 56932 edges. The original dataset comes from \url{https://dblp.org/xml/release}.

Note that, only Cora and DBLP have node labels. During the embedding phase, there is no testing set, but the learned node embeddings (not dedicated to a specific downstream task) will be evaluated by various downstream tasks. For the downstream tasks, there might be a training and/or testing set depending on the nature of a downstream task.

\subsubsection{Methods}
The proposed DNE method--GloDyNE is compared with the following state-of-the-art DNE methods for demonstrating the effectiveness and efficiency.

BCGD$^g$ \cite{zhu2016scalable}: The general objective of BCGD is to minimize the quadratic loss of reconstructing the network proximity matrix using the node embedding matrix with a temporal regularization term. BCGD$^g$ (or BCGD-global) employs all historical snapshots to jointly and cyclically update embeddings for all time steps.

BCGD$^l$ \cite{zhu2016scalable}: Unlike BCGD$^g$ but following the same general objective of BCGD as above, BCGD$^l$ (or BCGD-local) iteratively employs the previous snapshot and initializes current embeddings with the previous embeddings to update embeddings for a current time step.

DynGEM \cite{goyal2017dyngem}: This work proposes a strategy to modify the structure of a deep auto-encoder model based on the size of a current snapshot. At each time step, the auto-encoder model is initialized by its previous model. DynGEM continuously trains the adaptive auto-encoder model based on the existing edges in a current snapshot.

DynLINE \cite{du2018dynamic}: This work extends the static network embedding method--LINE\cite{tang2015line} to cope with dynamic networks. To improve efficiency, DynLINE updates the embeddings for the most affected nodes and new nodes in each snapshot.

DynTriad \cite{zhou2018dynamic}: DynTriad models the triadic closure process, social homophily, and temporal smoothness in its objective function to learn node embeddings at each time step. It optimizes the objective function according to the existing edges of each snapshot respectively.

tNE \cite{singer2019node}: tNE runs a static network embedding method to get node embeddings for each snapshot, and then exploits the temporal dependence among all available static node embeddings using Recurrent Neural Networks, so as to obtain the final node embeddings for a current time step.

The original open source codes with the default settings of BCGD\footnote{\url{https://github.com/linhongseba/Temporal-Network-Embedding}}, DynGEM\footnote{\url{https://github.com/palash1992/DynamicGEM}},  DynLINE\footnote{ \url{https://github.com/lundu28/DynamicNetworkEmbedding}}, DynTriad\footnote{\url{https://github.com/luckiezhou/DynamicTriad}}, and tNE\footnote{\url{https://github.com/urielsinger/tNodeEmbed}} are adopted in the experiments. Note that, BCGD$^g$ and BCGD$^l$ are two proposed algorithms in BCGD correspondingly to the type of algorithm 2 and 4. Moreover, we adopt the link prediction architecture of tNE to obtain node embeddings, so that all methods only use network linkage information as the supervised signal to learn node embeddings. Furthermore, the dimensionality of node embeddings is set to 128 for all methods for the fair comparison.

Regarding our method--GloDyNE, following \cite{perozzi2014deepwalk} and \cite{grover2016node2vec}, the hyper-parameters of walks per node, walk length, window size, and negative samples are set to 10, 80, 10, and 5 respectively. The hyper-parameter $\alpha$ to control the number of selected nodes for freely trade-off between effectiveness and efficiency, is set to 0.1 unless otherwise specified.

\subsection{Comparative Studies of Different Methods} \label{Sec5.2}
In this section, three typical types of downstream tasks are employed to evaluate the quality of obtained node embeddings by the seven methods on the six datasets. In particular, the \textit{graph reconstruction} task is used to demonstrate the ability of global topology preservation, while the \textit{link prediction} task and \textit{node classification} task are used to show the benefit of global topology preservation. For fairness, we first take out the node embeddings obtained by each method respectively, and then feed them to exactly the same downstream tasks. The above process is repeated for 20 runs. Their average results as well as other statistics are reported in Section \ref{GR}, \ref{LP}, and \ref{NC}. Moreover, the average results of the wall-clock time to obtain node embeddings, are reported in Section \ref{Time} for comparing the efficiency of the implementation of the seven methods. Finally, the overall performance regarding both effectiveness and efficiency is discussed in Section \ref{trade-off}.

All experiments in Section \ref{Sec5.2} are conducted in the following hardware specification. For all methods, we enable 32 Intel-Xeon-E5-2.2GHz logical CPUs and 512G memory. In addition, for DynGEM, DynTriad, DynLINE and tNE that can use GPU for acceleration, we also enable 1 Nvidia-Tesla-P100 GPU with 16G memory. The n/a values for DynLINE and tNE on AS733 are due to the inability of handling node deletions. The n/a values for DynGEM on HepPh and FBW are because of running out of GPU memory.

\subsubsection{Graph Reconstruction (GR)} \label{GR}
In order to demonstrate the ability of the global topology preservation of each method, one possible way is to use the obtained node embeddings to reconstruct the original network. For this purpose, precision at $k$ or $P@k$ is used as the metric to evaluate how well the top-$k$ similar nodes of each node in the embedding space can match the ground-truth neighbors of each node in the original network \cite{zhang2018arbitrary,zhu2018high,cao2015grarep}. Concretely, $P@k(v_i)=|\mathcal{Q}(v_i)_{@k} \cap \mathcal{N}(v_i)|~/~min(k, |\mathcal{N}(v_i)|)$, where $\mathcal{Q}(v_i)_{@k}$ gives a set of the top-$k$ similar nodes of a queried node $v_i$ based on the cosine similarity between node embeddings, and $\mathcal{N}(v_i)$ denotes a set of the ground-truth neighbors of $v_i$. As a result, the testing set is just a node set to query, but there is no training set. To show the ability of global topology preservation, we further calculate the mean of $P@k$ over all nodes in a current snapshot $G^t$, i.e., Mean$P@k=[~\sum\nolimits_{v_i^t \in \mathcal{V}^t} P@k(v^t_i)~]~/~|\mathcal{V}^t|$ where $\mathcal{V}^t$ is a set of all nodes in $G^t$, and $|\mathcal{V}^t|$ counts the number of nodes in $\mathcal{V}^t$. Table \ref{Tab1} presents the results for Mean$P@1$, Mean$P@5$, Mean$P@10$, Mean$P@20$, and Mean$P@40$.

\begin{table}[htbp]
  \centering
  \caption{Mean$P@k$ scores (in $\%$) of graph reconstruction tasks. Each entry of the table is obtained by the mean of Mean$P@k$ over all time steps, and then the mean with its standard deviation over 20 runs. Two-trailed and two-sample Student's T-Test is applied to the best two results (in bold). The best result is indicated by $^{\dagger}$ or $^{\ddagger}$, if the p-value is $<0.05$ or $<0.01$ compared to the second best result. The n/a values are due to the inability of handling node deletions or running out of memory.}
  \renewcommand\tabcolsep{1.65pt}
    \scalebox{0.8}{
    \begin{tabular}{l|llllll}
    \toprule
    \multicolumn{1}{c}{} & \multicolumn{1}{c}{AS733} & \multicolumn{1}{c}{Cora} & \multicolumn{1}{c}{DBLP} & \multicolumn{1}{c}{Elec} & \multicolumn{1}{c}{FBW} & \multicolumn{1}{c}{HepPh} \\
    \midrule
    \multicolumn{1}{c}{} & \multicolumn{6}{c}{Mean$P@1$} \\
    \midrule
    BCGD$^g$ & 02.13$\pm$0.06 & 02.09$\pm$0.16 & 02.17$\pm$0.08 & 10.01$\pm$0.05 & 00.32$\pm$0.01 & 32.81$\pm$0.07 \\
    BCGD$^l$ & 38.47$\pm$1.33 & 06.31$\pm$0.26 & 04.54$\pm$0.28 & 25.77$\pm$0.73 & 07.09$\pm$0.11 & 73.25$\pm$0.37 \\
    DynGEM & 00.89$\pm$0.03 & 11.26$\pm$0.55 & 30.47$\pm$0.60 & 04.41$\pm$0.06 & \multicolumn{1}{c}{n/a} & \multicolumn{1}{c}{n/a} \\
    DynLINE & \multicolumn{1}{c}{n/a} & 06.77$\pm$0.00 & 21.49$\pm$0.00 & 01.76$\pm$0.00 & 00.45$\pm$0.00 & 51.24$\pm$0.00 \\
    DynTriad & \textbf{65.43$\pm$18.76} & \textbf{64.68$\pm$23.85} & 68.91$\pm$21.42 & \textbf{69.86$\pm$8.48}$^{\ddagger}$ & \textbf{76.92$\pm$7.55} & \textbf{80.64$\pm$3.67} \\
    tNE   & \multicolumn{1}{c}{n/a} & 62.58$\pm$0.27 & \textbf{72.20$\pm$0.09} & 08.27$\pm$0.09 & 40.14$\pm$0.39 & 73.58$\pm$0.46 \\
    GloDyNE & \textbf{66.47$\pm$0.33} & \textbf{77.41$\pm$0.27}$^{\dagger}$ & \textbf{81.85$\pm$0.11}$^{\ddagger}$ & \textbf{51.75$\pm$0.16} & \textbf{90.63$\pm$0.04}$^{\ddagger}$ & \textbf{84.21$\pm$0.13}$^{\ddagger}$ \\
    \midrule
    \multicolumn{1}{c}{} & \multicolumn{6}{c}{Mean$P@5$} \\
    \midrule
    BCGD$^g$ & 02.00$\pm$0.05 & 10.23$\pm$0.32 & 00.68$\pm$0.03 & 10.41$\pm$0.04 & 00.15$\pm$0.00 & 31.08$\pm$0.04 \\
    BCGD$^l$ & 42.61$\pm$2.07 & 05.75$\pm$0.84 & 02.67$\pm$0.15 & 20.64$\pm$0.63 & 05.69$\pm$0.13 & 65.91$\pm$0.39 \\
    DynGEM & 00.82$\pm$0.02 & 07.22$\pm$0.36 & 22.62$\pm$0.48 & 04.14$\pm$0.04 & \multicolumn{1}{c}{n/a} & \multicolumn{1}{c}{n/a} \\
    DynLINE & \multicolumn{1}{c}{n/a} & 09.45$\pm$0.00 & 26.58$\pm$0.00 & 01.54$\pm$0.00 & 00.28$\pm$0.00 & 44.36$\pm$0.00 \\
    DynTriad & \textbf{62.18$\pm$17.47} & 53.40$\pm$23.13 & 56.15$\pm$22.02 & \textbf{66.64$\pm$9.13} & \textbf{58.84$\pm$11.09} & \textbf{73.96$\pm$4.35} \\
    tNE   & \multicolumn{1}{c}{n/a} & \textbf{60.81$\pm$0.34} & \textbf{68.16$\pm$0.19} & 06.82$\pm$0.06 & 27.54$\pm$0.34 & 61.24$\pm$0.44 \\
    GloDyNE & \textbf{70.37$\pm$0.26}$^{\dagger}$ & \textbf{79.66$\pm$0.08}$^{\ddagger}$ & \textbf{84.24$\pm$0.12}$^{\ddagger}$ & \textbf{66.42$\pm$0.11} & \textbf{87.81$\pm$0.02}$^{\ddagger}$ & \textbf{76.24$\pm$0.11}$^{\dagger}$ \\
    \midrule
    \multicolumn{1}{c}{} & \multicolumn{6}{c}{Mean$P@10$} \\
    \midrule
    BCGD$^g$ & 03.37$\pm$0.09 & 18.60$\pm$0.24 & 01.50$\pm$0.14 & 10.28$\pm$0.05 & 0.12$\pm$0.00 & 30.82$\pm$0.04 \\
    BCGD$^l$ & 51.94$\pm$2.82 & 08.04$\pm$2.70 & 03.30$\pm$0.16 & 19.22$\pm$0.64 & 4.67$\pm$0.09 & 61.12$\pm$0.38 \\
    DynGEM & 00.82$\pm$0.02 & 07.28$\pm$0.33 & 22.54$\pm$0.50 & 03.99$\pm$0.03 & \multicolumn{1}{c}{n/a} & \multicolumn{1}{c}{n/a} \\
    DynLINE & \multicolumn{1}{c}{n/a} & 12.80$\pm$0.00 & 33.76$\pm$0.00 & 01.39$\pm$0.00 & 00.23$\pm$0.00 & 40.34$\pm$0.00 \\
    DynTriad & \textbf{67.19$\pm$16.79} & 55.05$\pm$23.22 & 57.60$\pm$21.94 & \textbf{67.37$\pm$8.95} & \textbf{56.38$\pm$11.29} & \textbf{70.31$\pm$4.73} \\
    tNE   & \multicolumn{1}{c}{n/a} & \textbf{67.98$\pm$0.38} & \textbf{77.53$\pm$0.24} & 06.37$\pm$0.05 & 27.05$\pm$0.35 & 55.16$\pm$0.42 \\
    GloDyNE & \textbf{78.25$\pm$0.20}$^{\ddagger}$ & \textbf{86.53$\pm$0.11}$^{\ddagger}$ & \textbf{94.01$\pm$0.07}$^{\ddagger}$ & \textbf{71.19$\pm$0.09} & \textbf{89.22$\pm$0.01}$^{\ddagger}$ & \textbf{72.43$\pm$0.11} \\
    \midrule
    \multicolumn{1}{c}{} & \multicolumn{6}{c}{Mean$P@20$} \\
    \midrule
    BCGD$^g$ & 50.34$\pm$0.71 & 26.88$\pm$0.18 & 13.08$\pm$0.32 & 09.63$\pm$0.05 & 00.11$\pm$0.00 & 30.27$\pm$0.03 \\
    BCGD$^l$ & 69.62$\pm$4.34 & 14.21$\pm$4.00 & 12.97$\pm$0.2 & 19.44$\pm$0.71 & 03.96$\pm$0.06 & 56.76$\pm$0.36 \\
    DynGEM & 00.89$\pm$0.02 & 08.52$\pm$0.36 & 24.33$\pm$0.51 & 03.81$\pm$0.03 & \multicolumn{1}{c}{n/a} & \multicolumn{1}{c}{n/a} \\
    DynLINE & \multicolumn{1}{c}{n/a} & 17.53$\pm$0.00 & 39.39$\pm$0.00 & 01.30$\pm$0.00 & 00.20$\pm$0.00 & 36.85$\pm$0.00 \\
    DynTriad & \textbf{73.00$\pm$15.66} & 59.54$\pm$23.47 & 61.66$\pm$21.98 & \textbf{69.47$\pm$8.60} & \textbf{57.97$\pm$11.40} & \textbf{67.12$\pm$5.16} \\
    tNE   & \multicolumn{1}{c}{n/a} & \textbf{76.44$\pm$0.42} & \textbf{85.15$\pm$0.27} & 06.17$\pm$0.06 & 29.03$\pm$0.37 & 49.48$\pm$0.41 \\
    GloDyNE & \textbf{85.40$\pm$0.17}$^{\dagger}$ & \textbf{93.15$\pm$0.02}$^{\ddagger}$ & \textbf{98.84$\pm$0.02}$^{\ddagger}$ & \textbf{74.31$\pm$0.07}$^{\dagger}$ & \textbf{91.79$\pm$0.01}$^{\ddagger}$ & \textbf{69.91$\pm$0.10}$^{\dagger}$ \\
    \midrule
    \multicolumn{1}{c}{} & \multicolumn{6}{c}{Mean$P@40$} \\
    \midrule
    BCGD$^g$ & \textbf{89.60$\pm$0.74} & 36.96$\pm$0.16 & 35.89$\pm$0.29 & 08.85$\pm$0.05 & 00.12$\pm$0.00 & 29.06$\pm$0.03 \\
    BCGD$^l$ & 84.52$\pm$5.27 & 23.57$\pm$4.48 & 28.44$\pm$0.35 & 27.57$\pm$0.95 & 04.16$\pm$0.11 & 54.25$\pm$0.39 \\
    DynGEM & 01.20$\pm$0.05 & 10.61$\pm$0.42 & 26.77$\pm$0.53 & 03.67$\pm$0.04 & \multicolumn{1}{c}{n/a} & \multicolumn{1}{c}{n/a} \\
    DynLINE & \multicolumn{1}{c}{n/a} & 21.81$\pm$0.00 & 43.86$\pm$0.00 & 01.39$\pm$0.00 & 00.22$\pm$0.00 & 33.89$\pm$0.00 \\
    DynTriad & 78.90$\pm$14.01 & 64.95$\pm$23.12 & 66.33$\pm$21.67 & \textbf{72.80$\pm$8.03} & \textbf{61.92$\pm$11.53} & \textbf{65.42$\pm$5.61} \\
    tNE   & \multicolumn{1}{c}{n/a} & \textbf{81.27$\pm$0.41} & \textbf{88.57$\pm$0.30} & 06.52$\pm$0.07 & 32.61$\pm$0.40 & 44.58$\pm$0.38 \\
    GloDyNE & \textbf{90.87$\pm$0.13}$^{\ddagger}$ & \textbf{95.31$\pm$0.01}$^{\ddagger}$ & \textbf{99.85$\pm$0.00}$^{\ddagger}$ & \textbf{76.95$\pm$0.04}$^{\dagger}$ & \textbf{94.95$\pm$0.00}$^{\ddagger}$ & \textbf{69.81$\pm$0.08}$^{\ddagger}$ \\
    \bottomrule
    \end{tabular}%
    }
  \label{Tab1}%
\end{table}%

First, GloDyNE consistently outperforms all other methods on all datasets (28/30 cases), except that DynTriad outperforms GloDyNE on Elec dataset under Mean$P@1$ and Mean$P@5$ (2/30 cases). Second, although DynTriad can obtain the second best results in many cases, the standard deviation of DynTriad is always very high (often larger than $5.0\%$). In contrast, the performance of GloDyNE is very stable, since the standard deviation is always very low (often smaller than $0.3\%$). Third, GloDyNE significantly outperforms the second best method in 25/30 cases according to the statistical hypothesis testing\footnote{ Two-tailed and two-sample Student's T-Test is applied with the null hypothesis that there is no statistically significant difference of the mean over 20 runs between the two best results.}.

The main reason of such superiority of GloDyNE in the GR task is that GloDyNE is designed to better preserve the global topology of a dynamic network at each time step, while the GR task is also employed for demonstrating the ability of global topology preservation. 

\subsubsection{Link Prediction (LP)} \label{LP}
The (dynamic) LP task aims to predict future edges at time step $t+1$ using the obtained node embeddings at $t$. The testing edges include both added and deleted edges from $t$ to $t+1$, plus other edges randomly sampled from the snapshot at $t+1$ for balancing existent edges (or positive samples) and non-existent edges (or negative samples). The LP task is then evaluated by Area Under the ROC Curve (AUC) score based on the cosine similarity between node embeddings \cite{zhu2016scalable,Fu2019Learning,liao2018attributed}. Table \ref{Tab2} presents the AUC scores for LP tasks, and each entry of the table is obtained in the similar way as described in Table \ref{Tab1}.

\begin{table}[htbp]
  \centering
  \caption{AUC scores of (dynamic) link prediction tasks.}
  \renewcommand\tabcolsep{1.65pt}
    \scalebox{0.8}{
    \begin{tabular}{l|llllll}
    \toprule
          \multicolumn{1}{c}{} & \multicolumn{1}{c}{AS733} & \multicolumn{1}{c}{Cora} & \multicolumn{1}{c}{DBLP} & \multicolumn{1}{c}{Elec} & \multicolumn{1}{c}{FBW} & \multicolumn{1}{c}{HepPh} \\
    \midrule
    BCGD$^g$ & \textbf{69.64$\pm$0.64} & 67.92$\pm$0.95 & 66.55$\pm$0.85 & 81.22$\pm$0.91 & 82.90$\pm$0.29 & 77.06$\pm$0.15 \\
    BCGD$^l$ & 62.69$\pm$1.05 & \textbf{81.46$\pm$1.64} & \textbf{84.86$\pm$1.17}$^{\ddagger}$ & 86.61$\pm$3.72 & \textbf{83.83$\pm$1.34} & \textbf{88.16$\pm$1.93}$^{\ddagger}$ \\
    DynGEM & 61.60$\pm$1.28 & 56.69$\pm$1.42 & 60.90$\pm$0.81 & 60.71$\pm$1.56 & \multicolumn{1}{c}{n/a} & \multicolumn{1}{c}{n/a} \\
    DynLINE & \multicolumn{1}{c}{n/a} & 65.20$\pm$0.85 & 57.21$\pm$0.58 & 58.80$\pm$0.68 & 65.14$\pm$0.31 & 62.20$\pm$0.03 \\
    DynTriad & 64.36$\pm$3.45 & 65.86$\pm$6.37 & 58.78$\pm$2.66 & \textbf{94.25$\pm$1.50}$^{\ddagger}$ & 78.54$\pm$4.81 & 85.10$\pm$2.63 \\
    tNE   & \multicolumn{1}{c}{n/a} & 79.94$\pm$0.68 & 59.26$\pm$0.71 & 61.50$\pm$1.19 & 68.38$\pm$0.48 & 77.16$\pm$0.52 \\
    GloDyNE & \textbf{82.10$\pm$0.32}$^{\ddagger}$ & \textbf{93.43$\pm$0.36}$^{\ddagger}$ & \textbf{74.56$\pm$0.67} & \textbf{87.03$\pm$0.85} & \textbf{87.86$\pm$0.18}$^{\ddagger}$ & \textbf{85.88$\pm$0.05} \\
    \bottomrule
    \end{tabular}%
    }
  \label{Tab2}%
\end{table}%

According to the statistical hypothesis testing, GloDyNE significantly outperforms the second best method on AS733, Cora, and FBW by 12.64$\%$, 11.97$\%$ and 4.03$\%$ respectively (3/6 cases), while GloDyNE obtains the second best results on other three datasets (3/6 cases). Overall, GloDyNE is also a good method for the (dynamic) LP task on most datasets, thanks to the high-order proximities being used for better preserving the global topology \cite{cao2015grarep}. In fact, the high-order proximity between nodes is an important temporal feature for predicting future edges. For example, the triadic closure process which tries to predict the third edge among three nodes if there have already been two edges among them, as modelled in DynTriad \cite{zhou2018dynamic}, can be easily realized by considering the second-order proximity via setting $l\geq 3$ and $s\geq 3$ (see Section \ref{Step3} and \ref{Step4}). In the experiments, we set $l=80$ and $s=10$. As a result, much higher order proximities (up to 10$^{th}$ order according to $s$) are considered for better preserving the global topology, which therefore provides more advanced temporal features (analogous to triadic closure process) to improve the performance of GloDyNE in LP tasks on most datasets.

However, not all high-order proximities are helpful on all kinds of datasets. For example, we can observe from Table \ref{Tab1} and \ref{Tab2} that DynTriad, which mainly considers the second-order proximity due to modeling the triadic closure process, can obtain the best performance on Elec dataset; while GloDyNE, which considers more high-order proximities, obtains the second best performance. This observation might be caused by some special characteristics of Elec. It is worth noticing that Elec is an election network and might be quite different from other kinds of networks.

\subsubsection{Node Classification (NC)} \label{NC}
The NC task aims to infer the most likely label for the nodes without labels. Specifically, $50\%$, $70\%$, and $90\%$ nodes are randomly picked respectively to train a one-vs-rest logistic regression classifier based on their embeddings and labels. The left nodes respectively are treated as the testing set. Note that, only Cora and DBLP are employed in NC tasks, as other datasets do not have node labels. At each time step, the latest node embeddings are employed as the input features to logistic regression classifier. The prediction of the trained classifier over the testing set are evaluated by Micro-F1 and Macro-F1 \cite{perozzi2014deepwalk,grover2016node2vec,singer2019node} respectively. Table \ref{Tab3} presents the F1 scores for NC tasks, and each entry of the table is obtained in the similar way as described in Table \ref{Tab1}.

\begin{table}[htbp]
  \centering
  \caption{Micro-F1 and Macro-F1 scores of node classification tasks. Three different proportions of training set are evaluated respectively.}
  \renewcommand\tabcolsep{1.65pt}
    \scalebox{0.8}{
    \begin{tabular}{l|lll|lll}
    \toprule
          \multicolumn{1}{c}{} & \multicolumn{3}{c|}{Cora} & \multicolumn{3}{c}{DBLP} \\
          \multicolumn{1}{c}{} & \multicolumn{1}{c}{0.5} & \multicolumn{1}{c}{0.7} & \multicolumn{1}{c|}{0.9} & \multicolumn{1}{c}{0.5} & \multicolumn{1}{c}{0.7} & \multicolumn{1}{c}{0.9} \\
    \midrule
          \multicolumn{1}{c}{} & \multicolumn{6}{c}{Micro-F1} \\
    \midrule
    BCGD$^g$ & 32.28$\pm$0.29 & 33.43$\pm$0.49 & 33.93$\pm$0.74 & 56.00$\pm$0.17 & 56.15$\pm$0.15 & 56.34$\pm$0.16 \\
    BCGD$^l$ & 36.49$\pm$0.53 & 38.24$\pm$0.66 & 39.59$\pm$0.72 & 56.48$\pm$0.18 & 56.71$\pm$0.27 & 57.12$\pm$0.35 \\
    DynGEM & 36.92$\pm$0.69 & 38.95$\pm$0.60 & 41.03$\pm$0.85 & 55.22$\pm$0.04 & 55.37$\pm$0.08 & 55.76$\pm$0.13 \\
    DynLINE & 40.48$\pm$0.00 & 42.21$\pm$0.00 & 42.83$\pm$0.00 & 55.95$\pm$0.00 & 56.77$\pm$0.00 & 57.46$\pm$0.00 \\
    DynTriad & 36.54$\pm$3.61 & 37.67$\pm$3.74 & 38.57$\pm$3.70 & 55.60$\pm$0.55 & 56.22$\pm$0.46 & 56.86$\pm$0.65 \\
    tNE   & \textbf{65.37$\pm$0.29} & \textbf{67.12$\pm$0.29} & \textbf{67.97$\pm$0.49} & \textbf{63.19$\pm$0.16} & \textbf{63.90$\pm$0.20} & \textbf{64.27$\pm$0.36} \\
    GloDyNE & \textbf{74.20$\pm$0.30}$^{\ddagger}$ & \textbf{75.22$\pm$0.39}$^{\ddagger}$ & \textbf{75.54$\pm$0.58}$^{\ddagger}$ & \textbf{64.73$\pm$0.28}$^{\ddagger}$ & \textbf{65.17$\pm$0.30}$^{\ddagger}$ & \textbf{66.40$\pm$0.37}$^{\ddagger}$ \\
    \midrule
          \multicolumn{1}{c}{} & \multicolumn{6}{c}{Macro-F1} \\
    \midrule
    BCGD$^g$ & 08.30$\pm$0.20 & 08.68$\pm$0.20 & 08.58$\pm$0.38 & 10.24$\pm$0.14 & 10.26$\pm$0.20 & 10.19$\pm$0.31 \\
    BCGD$^l$ & 12.23$\pm$0.62 & 12.85$\pm$0.79 & 13.32$\pm$0.86 & 11.19$\pm$0.27 & 11.44$\pm$0.35 & 11.56$\pm$0.39 \\
    DynGEM & 09.91$\pm$0.56 & 10.79$\pm$0.50 & 11.33$\pm$0.63 & 08.25$\pm$0.06 & 08.41$\pm$0.11 & 08.63$\pm$0.16 \\
    DynLINE & 22.65$\pm$0.00 & 23.98$\pm$0.00 & 23.91$\pm$0.00 & 16.32$\pm$0.00 & 16.94$\pm$0.00 & 15.74$\pm$0.00 \\
    DynTriad & 16.92$\pm$4.01 & 17.71$\pm$4.30 & 17.64$\pm$4.37 & 12.81$\pm$0.74 & 13.06$\pm$0.81 & 13.05$\pm$1.10 \\
    tNE   & \textbf{50.03$\pm$0.31} & \textbf{52.45$\pm$0.39} & \textbf{51.41$\pm$0.96} & \textbf{25.60$\pm$0.40} & \textbf{27.02$\pm$0.35} & \textbf{26.02$\pm$0.84} \\
    GloDyNE & \textbf{61.20$\pm$0.66}$^{\ddagger}$ & \textbf{62.75$\pm$0.71}$^{\ddagger}$ & \textbf{62.01$\pm$1.00}$^{\ddagger}$ & \textbf{29.87$\pm$0.75}$^{\ddagger}$ & \textbf{30.28$\pm$0.81}$^{\ddagger}$ & \textbf{29.99$\pm$1.05}$^{\ddagger}$ \\
    \bottomrule
    \end{tabular}%
    }
  \label{Tab3}%
\end{table}%

According to the statistical hypothesis testing, GloDyNE significantly outperforms the second best method on both detests, which demonstrates the benefit of global topology preservation in NC tasks. Moreover, GloDyNE achieves better performance on Cora than DBLP. The reason is that Cora is a citation network where the label/field of nodes/papers contains less noise (the field of a journal or conference often remains the same), while DBLP is a co-author network where the label/field of nodes/authors contains more noise (the field of an author varies over time or an author with few papers is not accurate). The approach to construct dynamic networks and to generate node labels for Cora and DBLP (following tNE \cite{singer2019node}), are described in Section \ref{datasets}.

\subsubsection{Walk-Clock Time During Embedding} \label{Time}
To conduct the downstream tasks in Section \ref{GR}, \ref{LP}, and \ref{NC}, the common step is to first obtain node embeddings which serve as the \textit{low dimensional hidden features} to the downstream tasks. In this section, the wall-clock time of obtaining node embeddings (but not including downstream tasks) over all time steps are reported in Table \ref{Tab4}.

\begin{table}[htbp]
  \centering
  \caption{Wall-clock time (in seconds) of obtaining node embeddings (but not including downstream tasks) over all time steps. Each result is given by the sum of the wall-clock time over all time steps, and then the mean over 20 runs. The total number of nodes and edges of a dynamic network over all snapshots is also attached.}
  \renewcommand\tabcolsep{5pt}
    \scalebox{0.92}{
    \begin{tabular}{l|rrrrrr}
    \toprule
    \multicolumn{1}{c}{} & \multicolumn{1}{c}{AS733} & \multicolumn{1}{c}{Cora} & \multicolumn{1}{c}{DBLP} & \multicolumn{1}{c}{Elec} & \multicolumn{1}{c}{FBW} & \multicolumn{1}{c}{HepPh} \\
    \midrule
    BCGD$^g$ & 2987  & 4486  & 9277  & 6513  & 30063 & 24119 \\
    BCGD$^l$ & 597   & 1214  & 2543  & 1941  & 10272 & 12091 \\
    DynGEM & 2021  & 3336  & 8054  & 1577  & n/a   & n/a \\
    DynLINE & n/a   & 809   & 814   & 1577  & 1740  & 2095 \\
    DynTriad & 109   & 208   & 318   & 1879  & 3408  & 16893 \\
    tNE   & n/a   & 2728  & 4497  & 9989  & 79987 & 66679 \\
    GloDyNE & \textbf{64} & \textbf{106} & \textbf{186} & \textbf{203} & \textbf{943} & \textbf{908} \\
    \midrule
    \# of nodes & 45~k  & 66~k  & 108~k & 147~k & 902~k & 295~k \\
    \# of edges & 91~k  & 216~k & 233~k & 2095~k & 3703~k & 18491~k \\
    \bottomrule
    \end{tabular}%
    }
  \label{Tab4}%
\end{table}%

It is obvious that GloDyNE is the most efficient method among all methods on all datasets. In addition, the superiority of efficiency of GloDyNE grows, as the size of a dynamic network (given by the number of nodes or edges over all snapshots) grows. The reasons are as follows. First, GloDyNE is scalable regarding its time complexity, since there is no quadratic or higher term in $|\mathcal{V}|$ and $|\mathcal{E}|$ as analyzed in Section \ref{Sec4.3}. Second, the implementation of Step 4 of GloDyNE is highly parallelized and optimized.

To further test the scalability of GloDyNE on a very large-scale dataset, a hyperlink network from \url{http://konect.cc/networks/link-dynamic-dewiki} is employed. We follow the same approach as described in Section \ref{datasets} to generate its dynamic network. Specifically, the gap between snapshots is one calendar day; the recent 11 snapshots (03/Aug./2011-13/Aug./2011) are taken out; the initial snapshot has 2,161,514 nodes and 39,578,432 edges; the final snapshot has 2,165,677 nodes and 39,705,237 edges; and the total number of nodes and edges over all snapshots are 23,795,061 and 435,918,113. During the embedding phase, for the initial snapshot or time step (i.e., offline stage), the wall-clock time for Step 3 and Step 4 is 110698s and 12258s. After that, the averaged wall-clock time \textit{per snapshot} over other 10 snapshots (i.e., online stage) for Step 1-2, Step 3, and Step 4 are 2769s (for network partition and node selection), 12388s (for capturing topological changes), and 1255s (for updating node embeddings) respectively. We may ignore the offline stage as it is only conducted once at the beginning. During the online stage, the overall time to obtain embeddings for one snapshot is 16412s or 4.56h, which is acceptable for the scenario of daily updating embeddings in this very large-scale hyperlink network. Note that, one may further reduce the overall time by parallelizing random walks over multiprocessors in Step 3, so as to overcome the main bottleneck of the current implementation of GloDyNE.

\subsubsection{Effectiveness and Efficiency} \label{trade-off}
\begin{figure}[htbp]
    \centering
    \includegraphics[width=0.48\textwidth]{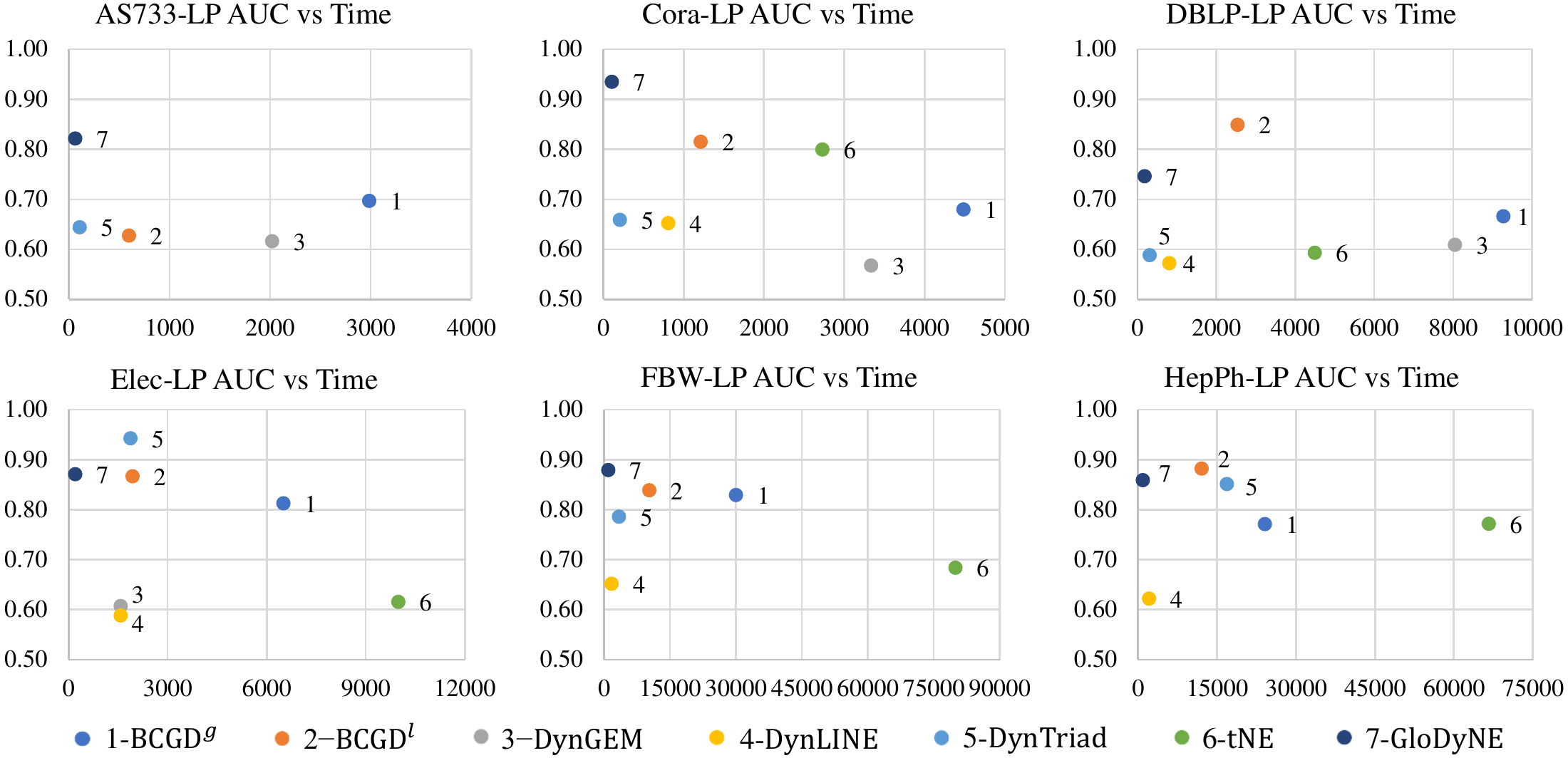}
    \caption{The comparison among the seven methods in terms of both effectiveness (y-axis for scores in decimal) and efficiency (x-axis for wall-clock time in seconds). We only show the plots of LP tasks for illustration. But the plots of GR and NC tasks are omitted, because GloDyNE obtains both the best effectiveness and best efficiency (i.e., located at the most top-left corner or the best choice) in almost all cases.}
    \label{Fig2}
\end{figure}

To better visualize the comparison among the seven methods in terms of both effectiveness and efficiency, we make scatter plots as shown in Figure \ref{Fig2} based on the quantitative results in above sections. Note that, the wall-clock time (of a method on a dataset) to obtain node embeddings in Section \ref{Time} is the same for different downstream tasks. Besides, the n/a values in the tables are omitted in Figure \ref{Fig2}.

In terms of both effectiveness and efficiency, GloDyNE is the best choice on AS733, Cora and FBW, since it is located at the most top-left corner in these cases. Although GloDyNE obtains the second best effectiveness on other three datasets, it still keeps the best efficiency in these cases. According to the quantitative results in Table \ref{Tab2} and Table \ref{Tab4}, from the \textit{effectiveness} perspective, GloDyNE is outperformed by BCGD$^l$, DynTriad and BCGD$^l$ by 10.3\%, 7,22\% and 2.28\% on DBLP, Elec and HepPh respectively. Nevertheless, from the \textit{efficiency} perspective, GloDyNE is $\times 11.5$, $\times 9.3$ and $\times 13.3$ faster respectively. Therefore, if one prefers efficiency, GloDyNE would be a better choice than BCGD$^l$, DynTriad and BCGD$^l$ on DBLP, Elec and HepPh respectively. 

Apart from the above discussions for LP tasks based on Figure \ref{Fig2}, the scatter plots of GR and NC tasks are omitted. The reason is that GloDyNE is the best choice in terms of both effectiveness and efficiency for 28/30 cases of GR tasks and 12/12 cases of NC tasks.

\subsection{Further Investigations of Proposed Method} \label{Sec5.3}
In this section, we further investigate the proposed method--GloDyNE. Since GloDyNE is proposed to better preserve the global topology, we focus on the ability of global topology preservation, and thus adopt the \textit{graph reconstruction} task for discussions. Besides, thanks to the good time and space efficiency of GloDyNE and its variants, all experiments in this section are conducted with less expensive hardware: 16 Intel-Xeon-E5-2.2GHz logical CPUs and 8G memory. All experiments are repeated for 20 runs. 
Due to the space limit of the paper, we choose two datasets for illustration. One is AS733 that includes both node additions and deletions, while another is Elec that only includes node additions.

\subsubsection{Necessity of Dynamic Network Embedding}
One advantage of DNE is that, it promptly updates node embeddings at each time step, so that the latest embeddings can better reflect the original network topology at each time step. To demonstrate this, two variants of GloDyNE based on the SGNS model, namely SGNS-static and SGNS-retrain, are used for comparison. For SGNS-static, we perform the $t=0$ part of Algorithm \ref{alg1}, and the obtained embeddings at $t=0$ will be identically used in the downstream task at each time step. For SGNS-retrain, we repeatedly perform the $t=0$ part of Algorithm \ref{alg1} at each time step, and the obtain embeddings at each time step will be used in the downstream task at each time step respectively.

\begin{figure}[htbp]
    \centering
    \includegraphics[width=0.48\textwidth]{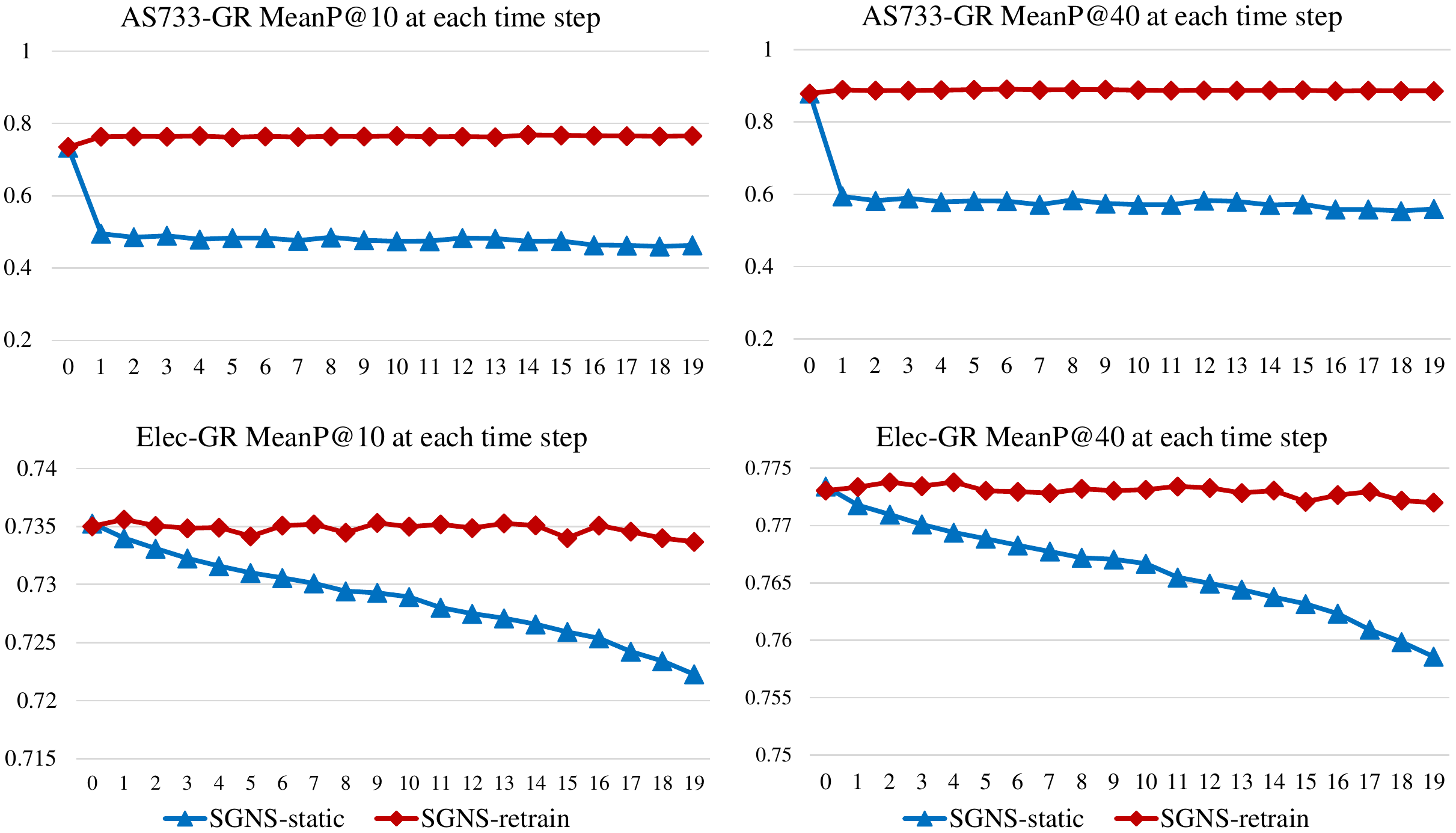}
    \caption{SGNS-static vs SGNS-retrain in graph reconstruct tasks for showing the necessity of dynamic network embedding: y-axis indicates Mean$P@k$ scores; x-axis indicates time steps; each point depicts the average result over 20 runs at a time step.}
    \label{Fig3}
\end{figure}

According to Figure \ref{Fig3}, SGNS-retrain outperforms SGNS-static on both datasets. For AS733, SGNS-retrain maintains the performance at a superior level all the time, whereas the performance of SGNS-static suddenly decreases at $t=1$ and then maintains a poor level afterward. For Elec, SGNS-retrain maintains the performance at a superior level all the time, whereas the performance of SGNS-static gradually decreases. The difference of sudden drops on AS733 and gradual drops on Elec is due to the fact that the network topology between consecutive time steps on AS733 varies more severely than on Elec (see Section \ref{datasets}), so that the obtained node embeddings at $t=0$ is less useful afterward. Consequently, it is needed to promptly update node embeddings at each time step (i.e., the necessity of DNE) as what SGNS-retrain--the naive DNE method does.

\subsubsection{Incremental Learning vs Retraining}
Instead of SGNS-retrain, recent DNE methods often adopt the incremental learning paradigm by continuously training the previous model on a new training set. Another baseline--SGNS-increment thus follows Algorithm \ref{alg1} but replaces all operations in lines 7-14 with $\mathcal{V}^t_{sel}=\mathcal{V}^t_{all}$. The difference between SGNS-increment and SGNS-retrain is if they reuse the previous model as the initialization of next model.

\begin{figure}[htbp]
    \centering
    \includegraphics[width=0.48\textwidth]{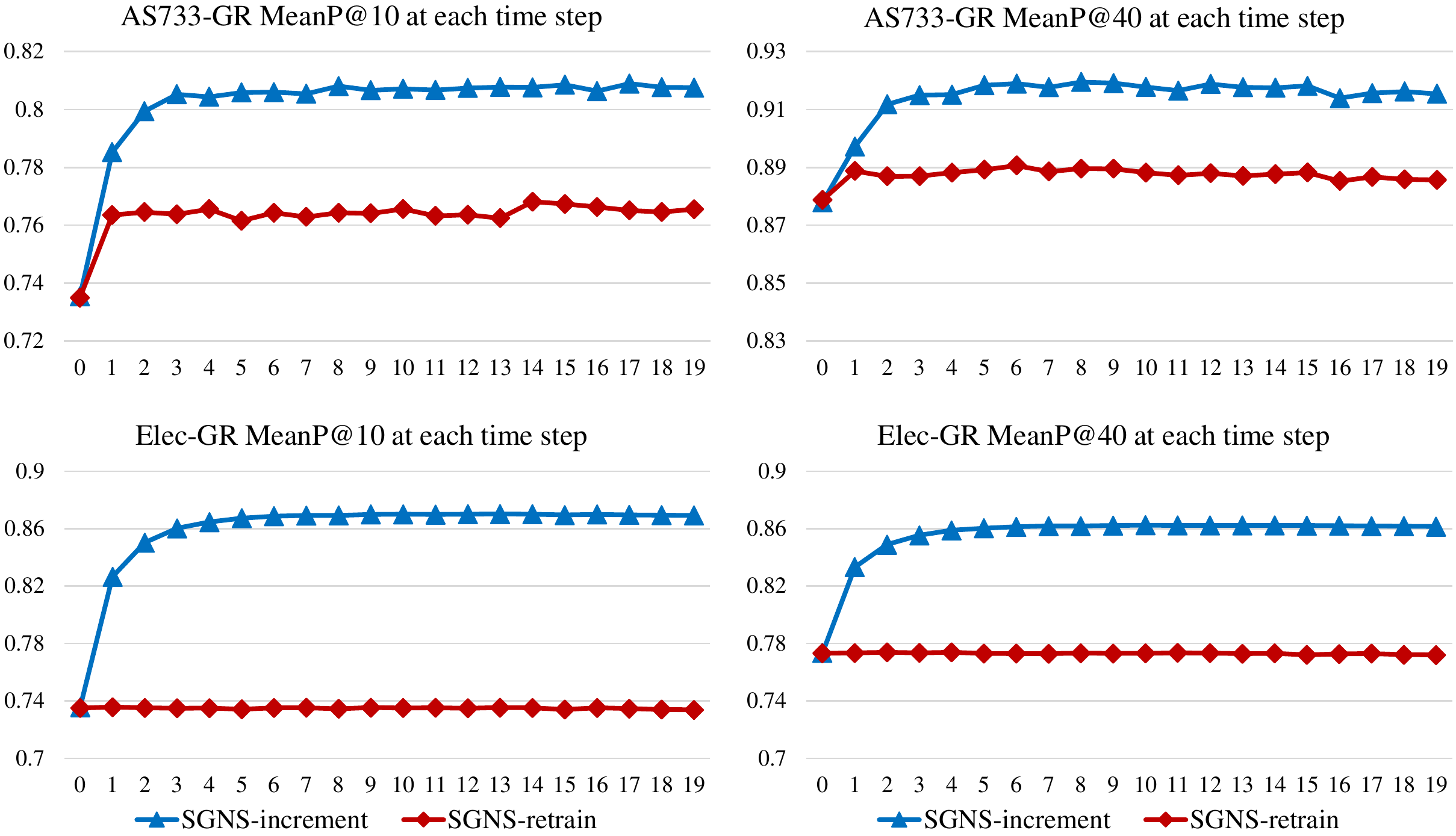}
    \caption{SGNS-increment vs SGNS-retrain in graph reconstruct tasks for showing the advantage of reusing previous models: y-axis indicates Mean$P@k$ scores; x-axis indicates time steps; each point depicts the average result over 20 runs at a time step.}
    \label{Fig4}
\end{figure}

According to Figure \ref{Fig4}, SGNS-increment outperforms SGNS-retrain on both datasets. The general tendency on both datasets is the same, although the performances of SGNS-increment and SGNS-retrain are both less stable on AS733 than on Elec, due to the larger variations between consecutive snapshots on AS733 (see Section \ref{datasets}). These observations show that reusing the previous model as the initialization of next model might be not only useful for a dynamic network with small variations, but also useful for a dynamic network with relative large variations.

\subsubsection{Visualization of Embeddings}
\begin{figure}[htbp]
    \centering
    \includegraphics[width=0.48\textwidth]{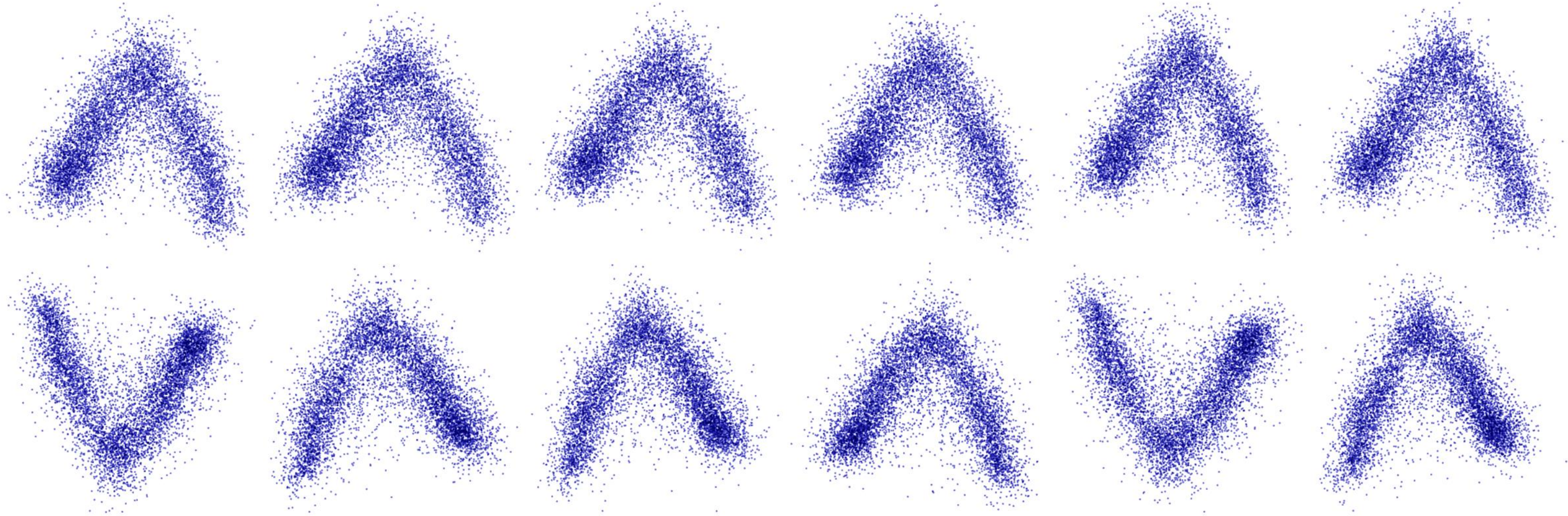}
    \caption{The first row of six sub-figures is for applying GloDyNE on Elec to obtain node embeddings for six consecutive time steps from 8 to 13 respectively. Similarly, the second row is for SGNS-retrain. To visualize embeddings, we further project them from 128 to 2 dimensions.}
    \label{Fig5}
\end{figure}

To show how embeddings evolve over consecutive time steps, we employ Principle Component Analysis to project the obtained node embeddings from 128 to 2 dimensions. As shown in Figure \ref{Fig5}, GloDyNE keeps not only the relative position but also the absolute position of node embeddings between two consecutive time steps, whereas SGNS-retrain cannot keep the absolute position (notice the rotation of the 'v' shape). The reason of why GloDyNE can well keep the absolute position is owing to the incremental learning paradigm which acts as an implicit smoothing mechanism.

\subsubsection{Different Node Selecting Strategies}
According to Figure \ref{Fig3} and \ref{Fig4}, the ranking among the three baselines is SGNS-increment$>$SGNS-retrain$>$SGNS-static. Although SGNS-increment (i.e., GloDyNE with $\alpha = 1.0$) achieves the best performance, it is not efficient enough since all nodes in a current snapshot are selected for conducting random walks and then training the SGNS model. One natural idea for further improving the efficiency is to select some representative nodes as the \textit{approximate solution}, such that it can significantly reduce the wall-clock time but meanwhile, still retain a good performance. Consequently, in this work, we propose a node selecting strategy, denoted as $S4$, as described in Section \ref{Step1} and Section \ref{Step2}.

\begin{table}[htbp]
  \centering
  \caption{The performance of GloDyNE with different node selecting strategies w.r.t. different length of random walks in graph reconstruction tasks. Each entry is obtained in the similar way as described in Table \ref{Tab1}.}
  \renewcommand\tabcolsep{4pt}
    \scalebox{0.88}{
    \begin{tabular}{c|llll|llll}
    \toprule
    \multicolumn{1}{c}{} & \multicolumn{4}{c|}{AS733}    & \multicolumn{4}{c}{Elec} \\
    \midrule
          & \multicolumn{1}{c}{$S1$} & \multicolumn{1}{c}{$S2$} & \multicolumn{1}{c}{$S3$} & \multicolumn{1}{c|}{$S4$} & \multicolumn{1}{c}{$S1$} & \multicolumn{1}{c}{$S2$} & \multicolumn{1}{c}{$S3$} & \multicolumn{1}{c}{$S4$} \\
    \midrule
    $l$     & \multicolumn{8}{c}{Mean$P@10$} \\
    \midrule
    3     & 15.275 & 18.841 & 20.623 & \textbf{21.340}$^{\ddagger}$ & 04.038 & 05.912 & 06.105 & \textbf{06.122} \\
    5     & 36.791 & 38.657 & 39.671 & \textbf{40.033}$^{\ddagger}$ & 07.091 & 11.204 & 11.519 & \textbf{11.837}$^{\ddagger}$ \\
    8     & 43.976 & 44.289 & 44.807 & \textbf{44.862} & 10.518 & 14.886 & 15.357 & \textbf{15.630}$^{\ddagger}$ \\
    10    & 44.714 & 44.934 & \textbf{45.253} & 45.225 & 12.029 & 17.446 & 18.059 & \textbf{18.325}$^{\ddagger}$ \\
    15    & 48.083 & 48.311 & 48.495 & \textbf{48.514} & 17.704 & 25.468 & 25.963 & \textbf{26.420}$^{\ddagger}$ \\
    20    & 53.713 & 54.083 & 54.323 & \textbf{54.443} & 26.656 & 34.437 & 34.790 & \textbf{35.071}$^{\ddagger}$ \\
    30    & 63.585 & 63.881 & 64.163 & \textbf{64.186} & 44.556 & 48.710 & 48.938 & \textbf{49.074}$^{\dagger}$ \\
    40    & 69.483 & 69.694 & \textbf{69.961} & 69.852 & 54.906 & 57.055 & 57.157 & \textbf{57.220} \\
    50    & 72.918 & 73.083 & \textbf{73.324} & 73.281 & 60.846 & 62.224 & 62.278 & \textbf{62.306} \\
    60    & 75.086 & 75.265 & \textbf{75.507} & 75.497 & 65.021 & 65.981 & \textbf{66.031} & 66.028 \\
    70    & 76.491 & 76.724 & 76.985 & \textbf{77.046} & 68.288 & 68.946 & 68.920 & \textbf{69.004}$^{\dagger}$ \\
    80    & 77.723 & 77.982 & \textbf{78.385} & 78.208 & 70.825 & 71.272 & 71.332 & \textbf{71.340} \\
    90    & 78.778 & 79.090 & 79.367 & \textbf{79.369} & 72.916 & 73.257 & 73.268 & \textbf{73.277} \\
    100   & 79.846 & 79.991 & 80.227 & \textbf{80.305} & 74.621 & 74.882 & 74.870 & \textbf{74.885} \\
    \midrule
    $l$     & \multicolumn{8}{c}{Mean$P@40$} \\
    \midrule
    3     & 22.097 & 26.784 & 28.791 & \textbf{29.747}$^{\ddagger}$ & 04.972 & 06.587 & 06.820 & \textbf{06.822} \\
    5     & 52.509 & 54.063 & 54.695 & \textbf{55.094}$^{\ddagger}$ & 05.925 & 11.482 & 11.936 & \textbf{12.312}$^{\ddagger}$ \\
    8     & 59.334 & 59.606 & 59.809 & \textbf{59.845} & 07.532 & 14.409 & 15.183 & \textbf{15.608}$^{\ddagger}$ \\
    10    & 60.522 & 60.871 & 61.005 & \textbf{61.062} & 09.278 & 18.567 & 19.401 & \textbf{19.887}$^{\ddagger}$ \\
    15    & 66.123 & 66.801 & \textbf{67.372} & 67.340 & 20.117 & 32.013 & 32.617 & \textbf{33.224}$^{\ddagger}$ \\
    20    & 72.998 & 73.506 & 74.077 & \textbf{74.095} & 35.458 & 44.739 & 45.027 & \textbf{45.251}$^{\ddagger}$ \\
    30    & 81.225 & 81.617 & 82.111 & \textbf{82.191} & 57.477 & 60.500 & 60.604 & \textbf{60.739}$^{\ddagger}$ \\
    40    & 85.083 & 85.506 & 86.007 & \textbf{86.018} & 66.260 & 67.436 & 67.488 & \textbf{67.502} \\
    50    & 87.332 & 87.705 & \textbf{88.109} & 88.105 & 70.602 & 71.218 & 71.209 & \textbf{71.215} \\
    60    & 88.756 & 89.102 & 89.327 & \textbf{89.329} & 73.414 & 73.731 & \textbf{73.725} & 73.714 \\
    70    & 89.713 & 89.913 & 90.085 & \textbf{90.156} & 75.493 & 75.584 & 75.564 & \textbf{75.576} \\
    80    & 90.474 & 90.648 & \textbf{90.779} & 90.695 & 77.046 & 77.014 & 77.016 & \textbf{77.027}\\
    90    & 91.103 & 91.217 & 91.259 & \textbf{91.286} & 78.300 & 78.214 & 78.179 & \textbf{78.189} \\
    100   & 91.667 & 91.608 & 91.703 & \textbf{91.745} & 79.287 & 79.150 & \textbf{79.127} & 79.126 \\
    \bottomrule
    \end{tabular}%
    }
  \label{Tab5}%
\end{table}%

In order to show the advantage of $S4$ used in GloDyNE, the following baselines with different node selecting strategies are used for comparison. For fairness, the number of selected nodes at each time step is set to $\alpha |\mathcal{V}^t|=0.1|\mathcal{V}^t|$ for all strategies. Concretely, $S1$ selects the nodes randomly with replacement from the reservoir $\mathcal{R}^t$ which records the most affected nodes; $S2$ selects the nodes randomly without replacement from $\mathcal{R}^t$ and then from all nodes in a current snapshot if $|\mathcal{R}^t|<0.1|\mathcal{V}^t|$; $S3$ selects the nodes randomly without replacement from all nodes in a current snapshot. Intuitively, from the perspective of \textit{diversity of selected nodes}, $S1<S2<S3<S4$ because 1) sampling nodes from $\mathcal{R}^t$ cannot be aware of inactive sub-networks which exist in many real-world dynamic networks; 2) sampling nodes from all nodes in a current snapshot cannot guarantee the selected nodes have an enough distance from each other; 3) sampling one node from each sub-network after network partition as introduced in $S4$, however, can ensure the selected nodes have an enough distance from each other.

To compare the performance of GloDyNE with different node selecting strategies, the length of random walks $l$ (see Section \ref{Step3}) should be also considered. Because as $l$ increases, the generated random walks (or node sequences) become less distinguishable. An extreme case is that, if $l$ goes to infinity, a random walker starting from any node in a network can well explore its global topology. As a result, we compare the four different node selecting strategies w.r.t. different $l$s as shown in Table \ref{Tab5}.

According to Table \ref{Tab5}, first, the overall ranking of the performance under a same $l$ is $S1<S2<S3<S4$, which exactly matches the ranking of the diversity of selected nodes as discussed above. Second, as $l$ increases, the four strategies become less distinguishable, which verifies the above analysis of four strategies w.r.t. $l$. Third, comparing AS733 and Elec, it shows that the superiority of $S4$ over other three node selecting strategies, is more obvious on Elec than on AS733. It suggests that using $S4$ with GloDyNE on a lager dataset (e.g., Elec is larger than AS733 as shown in Table \ref{Tab4}) might gain more benefits.

\subsubsection{The Free Hyper-Parameter}
\begin{figure}[htbp]
    \centering
    \includegraphics[width=0.48\textwidth]{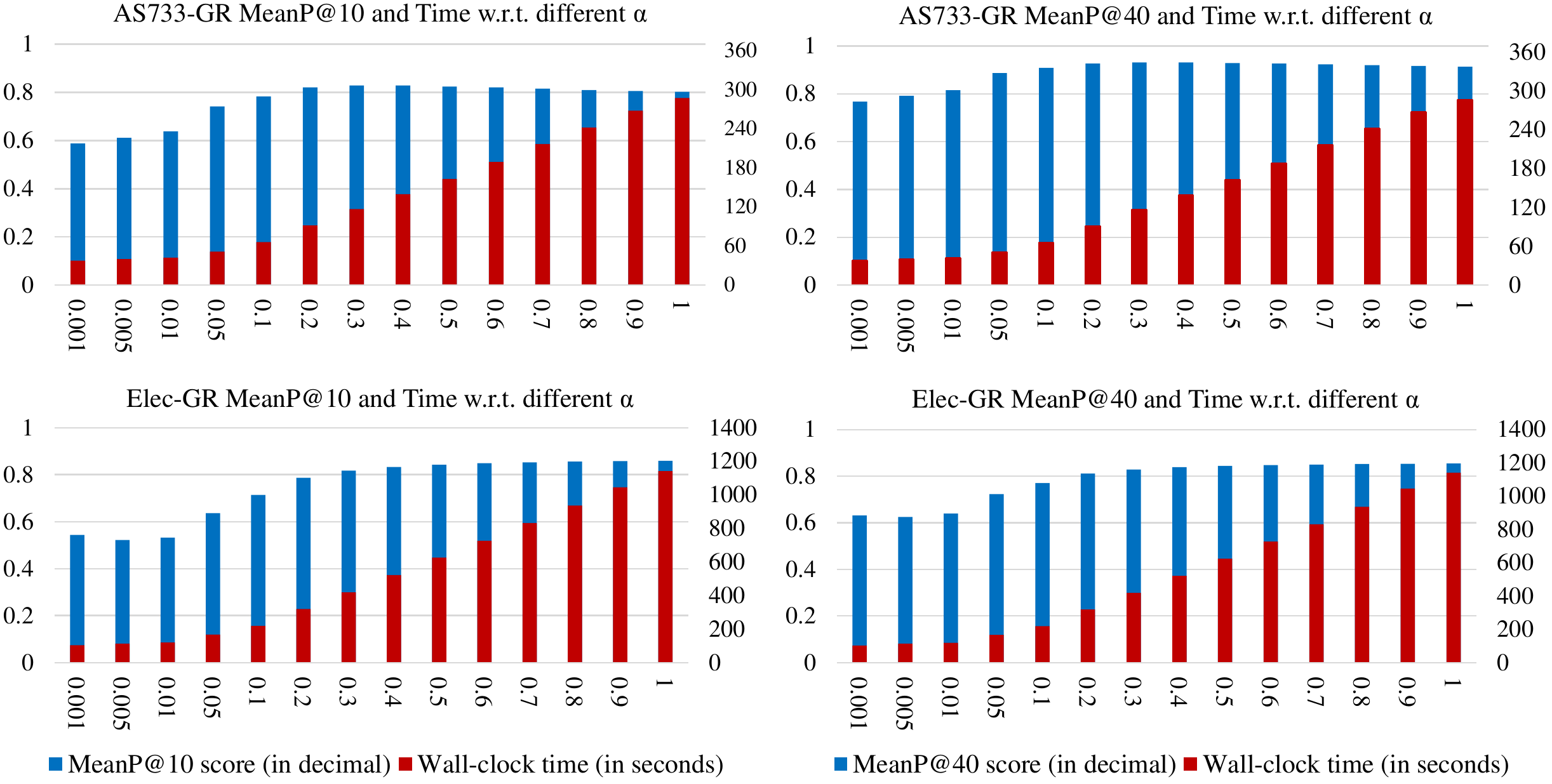}
    \caption{The effectiveness (in blue corresponding to the left y-axis) and efficiency (in red corresponding to the right y-axis) of GloDyNE w.r.t. different $\alpha$ (x-axis) which determines the number of selected nodes.}
    \label{Fig6}
\end{figure}

The hyper-parameter $\alpha$, which determines the number of selected nodes at each time step, is designed for freely trade-off between effectiveness and efficiency. We vary $\alpha$ from 0.1 to 1.0 with step 0.1, together with other four smaller values. Each bar in Figure \ref{Fig6} has two results. The blue one shows the effectiveness which is measured by the mean of Mean$P@k$ over all time steps and over 20 runs, while the red one shows the efficiency which is measured by the mean over 20 runs of the total wall-clock time over all time steps.

According to Figure \ref{Fig6}, it demonstrates that the hyper-parameter $\alpha$ can be used to freely compromise between effectiveness and efficiency. With this free hyper-parameter, one could fulfill the real-world requirement by trade-off between effectiveness and efficiency, if downstream tasks require the latest node embeddings within a specified period. Besides, all experiments in the above sections set $\alpha = 0.1$, which implies one can obtain better results by increasing $\alpha$ at the risk of consuming more wall-clock time.

Furthermore, an interesting observation is that increasing $\alpha$ to a certain level achieves a very competitive performance as $\alpha=1.0$ (GloDyNE with $\alpha=1.0$ is equivalent to SGNS-increment), but consumes much less wall-clock time. This observation also supports that GloDyNE especially the proposed node selecting strategy that selects partial nodes, makes a good approximation to SGNS-increment that selects all nodes for further computation.

\section{Conclusion} \label{Sec6}
This work proposed a new DNE method--GloDyNE, which aims to efficiently update node embeddings while better preserving the global topology of a dynamic network at each time step, by extending the SGNS model to an incremental learning paradigm. In particular, unlike all previous DNE methods, a novel node selecting strategy is proposed to diversely select the representative nodes over a network, so as to additionally considers the inactive sub-networks for better global topology preservation. The extensive experiments not only confirmed the effectiveness and efficiency of GloDyNE w.r.t. other six state-of-the-art DNE methods, but also verified the usefulness of some special designs or considerations in the proposed method.

From a high-level view, GloDyNE can also be seen as a general DNE framework based on the incremental learning paradigm of SGNS model. With this framework, one may design a different node selecting strategy to preserve other desirable topological features into node embeddings for a specific application. On the other hand, the idea of selecting diverse nodes could be adapted to other existing DNE methods for better global topology preservation. Besides, one more future work, according to Figure \ref{Fig6}, is to further investigate why selecting partial nodes can receive almost the same performance or even the superior performance compared to selecting all nodes.

\appendices


\ifCLASSOPTIONcompsoc
  \section*{Acknowledgments}
\else
  \section*{Acknowledgment}
\fi
The authors would like to thank the anonymous reviewers for their constructive comments. This work was supported in part by the National Key Research and Development Program of China under Grant 2017YFB1003102, in part by the Guangdong Provincial Key Laboratory under Grant 2020B121201001, in part by the Natural Science Foundation of China under Grant 61672478, in part by the Program for Guangdong Introducing Innovative and Entrepreneurial Teams under Grant 2017ZT07X386, in part by the Shenzhen Peacock Plan under Grant KQTD2016112514355531, and in part by the National Leading Youth Talent Support Program of China.

\ifCLASSOPTIONcaptionsoff
  \newpage
\fi



\bibliographystyle{IEEEtran}
\bibliography{IEEE_TKDE.bbl}

\begin{thebibliography}{10}
\providecommand{\url}[1]{#1}
\csname url@samestyle\endcsname
\providecommand{\newblock}{\relax}
\providecommand{\bibinfo}[2]{#2}
\providecommand{\BIBentrySTDinterwordspacing}{\spaceskip=0pt\relax}
\providecommand{\BIBentryALTinterwordstretchfactor}{4}
\providecommand{\BIBentryALTinterwordspacing}{\spaceskip=\fontdimen2\font plus
\BIBentryALTinterwordstretchfactor\fontdimen3\font minus
  \fontdimen4\font\relax}
\providecommand{\BIBforeignlanguage}[2]{{%
\expandafter\ifx\csname l@#1\endcsname\relax
\typeout{** WARNING: IEEEtran.bst: No hyphenation pattern has been}%
\typeout{** loaded for the language `#1'. Using the pattern for}%
\typeout{** the default language instead.}%
\else
\language=\csname l@#1\endcsname
\fi
#2}}
\providecommand{\BIBdecl}{\relax}
\BIBdecl

\bibitem{cui2018survey}
P.~Cui, X.~Wang, J.~Pei, and W.~Zhu, ``A survey on network embedding,''
  \emph{{IEEE} Trans. Knowl. Data Eng.}, vol.~31, no.~5, pp. 833--852, 2018.

\bibitem{hamilton2017representation}
W.~L. Hamilton, R.~Ying, and J.~Leskovec, ``Representation learning on graphs:
  Methods and applications,'' \emph{{IEEE} Data Eng. Bull.}, vol.~40, no.~3,
  pp. 52--74, 2017.

\bibitem{goyal2018graph}
P.~Goyal and E.~Ferrara, ``Graph embedding techniques, applications, and
  performance: A survey,'' \emph{Knowledge-Based Systems}, vol. 151, pp.
  78--94, 2018.

\bibitem{perozzi2014deepwalk}
B.~Perozzi, R.~Al{-}Rfou, and S.~Skiena, ``Deepwalk: online learning of social
  representations,'' in \emph{ACM SIGKDD Conference on Knowledge Discovery and
  Data Mining (KDD)}, 2014, pp. 701--710.

\bibitem{tang2015line}
J.~Tang, M.~Qu, M.~Wang, M.~Zhang, J.~Yan, and Q.~Mei, ``{LINE:} large-scale
  information network embedding,'' in \emph{International Conference on World
  Wide Web (WWW)}, 2015, pp. 1067--1077.

\bibitem{cao2015grarep}
S.~Cao, W.~Lu, and Q.~Xu, ``Grarep: Learning graph representations with global
  structural information,'' in \emph{Proceedings of the 24th ACM international
  on conference on information and knowledge management}.\hskip 1em plus 0.5em
  minus 0.4em\relax ACM, 2015, pp. 891--900.

\bibitem{grover2016node2vec}
A.~Grover and J.~Leskovec, ``node2vec: Scalable feature learning for
  networks,'' in \emph{ACM SIGKDD Conference on Knowledge Discovery and Data
  Mining (KDD)}, 2016, pp. 855--864.

\bibitem{ou2016asymmetric}
M.~Ou, P.~Cui, J.~Pei, Z.~Zhang, and W.~Zhu, ``Asymmetric transitivity
  preserving graph embedding,'' in \emph{ACM SIGKDD Conference on Knowledge
  Discovery and Data Mining (KDD)}, 2016, pp. 1105--1114.

\bibitem{zhu2016scalable}
L.~Zhu, D.~Guo, J.~Yin, G.~V. Steeg, and A.~Galstyan, ``Scalable temporal
  latent space inference for link prediction in dynamic social networks,''
  \emph{{IEEE} Trans. Knowl. Data Eng.}, vol.~28, no.~10, pp. 2765--2777, 2016.

\bibitem{li2017attributed}
J.~Li, H.~Dani, X.~Hu, J.~Tang, Y.~Chang, and H.~Liu, ``Attributed network
  embedding for learning in a dynamic environment,'' in \emph{Proceedings of
  the 2017 ACM on Conference on Information and Knowledge Management}, 2017.

\bibitem{goyal2017dyngem}
P.~Goyal, N.~Kamra, X.~He, and Y.~Liu, ``Dyngem: Deep embedding method for
  dynamic graphs,'' in \emph{IJCAI International Workshop on Representation
  Learning for Graphs}, 2017.

\bibitem{zhu2018high}
D.~Zhu, P.~Cui, Z.~Zhang, J.~Pei, and W.~Zhu, ``High-order proximity preserved
  embedding for dynamic networks,'' \emph{{IEEE} Trans. Knowl. Data Eng.},
  vol.~30, no.~11, pp. 2134--2144, 2018.

\bibitem{zhang2018timers}
Z.~Zhang, P.~Cui, J.~Pei, X.~Wang, and W.~Zhu, ``{TIMERS:} error-bounded {SVD}
  restart on dynamic networks,'' in \emph{Proceedings of the Thirty-Second
  {AAAI} Conference on Artificial Intelligence (AAAI)}, 2018.

\bibitem{du2018dynamic}
L.~Du, Y.~Wang, G.~Song, Z.~Lu, and J.~Wang, ``Dynamic network embedding : An
  extended approach for skip-gram based network embedding,'' in
  \emph{Proceedings of the Twenty-Seventh International Joint Conference on
  Artificial Intelligence (IJCAI)}, 2018.

\bibitem{zhou2018dynamic}
L.~Zhou, Y.~Yang, X.~Ren, F.~Wu, and Y.~Zhuang, ``Dynamic network embedding by
  modeling triadic closure process,'' in \emph{Proceedings of the Thirty-Second
  {AAAI} Conference on Artificial Intelligence (AAAI)}, 2018.

\bibitem{chen2018scalable}
X.~Chen, P.~Cui, L.~Yi, and S.~Yang, ``Scalable optimization for embedding
  highly-dynamic and recency-sensitive data,'' in \emph{Proceedings of the 24th
  ACM SIGKDD International Conference on Knowledge Discovery \& Data
  Mining}.\hskip 1em plus 0.5em minus 0.4em\relax ACM, 2018, pp. 130--138.

\bibitem{mahdavi2018dynnode2vec}
S.~Mahdavi, S.~Khoshraftar, and A.~An, ``dynnode2vec: Scalable dynamic network
  embedding,'' in \emph{{IEEE} International Conference on Big Data (Big
  Data)}, 2018, pp. 3762--3765.

\bibitem{singer2019node}
U.~Singer, I.~Guy, and K.~Radinsky, ``Node embedding over temporal graphs,'' in
  \emph{Proceedings of the Twenty-Eighth International Joint Conference on
  Artificial Intelligence (IJCAI)}, 2019.

\bibitem{trivedi2019dyrep}
R.~Trivedi, M.~Farajtabar, P.~Biswal, and H.~Zha, ``Dyrep: Learning
  representations over dynamic graphs,'' in \emph{International Conference on
  Learning Representations}, 2019.

\bibitem{karypis1998fast}
G.~Karypis and V.~Kumar, ``A fast and high quality multilevel scheme for
  partitioning irregular graphs,'' \emph{SIAM Journal on scientific Computing},
  vol.~20, no.~1, pp. 359--392, 1998.

\bibitem{mikolov2013distributed}
T.~Mikolov, I.~Sutskever, K.~Chen, G.~S. Corrado, and J.~Dean, ``Distributed
  representations of words and phrases and their compositionality,'' in
  \emph{Advances in Neural Information Processing Systems (NIPS)}, 2013, pp.
  3111--3119.

\bibitem{yu2018netwalk}
W.~Yu, W.~Cheng, C.~C. Aggarwal, K.~Zhang, H.~Chen, and W.~Wang, ``Netwalk: {A}
  flexible deep embedding approach for anomaly detection in dynamic networks,''
  in \emph{ACM SIGKDD Conference on Knowledge Discovery and Data Mining (KDD)},
  2018.

\bibitem{yang2015network}
C.~Yang, Z.~Liu, D.~Zhao, M.~Sun, and E.~Y. Chang, ``Network representation
  learning with rich text information.'' in \emph{IJCAI}, vol. 2015, 2015, pp.
  2111--2117.

\bibitem{gao2018deep}
H.~Gao and H.~Huang, ``Deep attributed network embedding.'' in \emph{IJCAI},
  vol.~18.\hskip 1em plus 0.5em minus 0.4em\relax New York, NY, 2018, pp.
  3364--3370.

\bibitem{huang2017accelerated}
X.~Huang, J.~Li, and X.~Hu, ``Accelerated attributed network embedding,'' in
  \emph{Proceedings of the 2017 SIAM international conference on data
  mining}.\hskip 1em plus 0.5em minus 0.4em\relax SIAM, 2017, pp. 633--641.

\bibitem{bulucc2016recent}
A.~Bulu{\c{c}}, H.~Meyerhenke, I.~Safro, P.~Sanders, and C.~Schulz, ``Recent
  advances in graph partitioning,'' in \emph{Algorithm Engineering}.\hskip 1em
  plus 0.5em minus 0.4em\relax Springer, 2016, pp. 117--158.

\bibitem{zhang2018arbitrary}
Z.~Zhang, P.~Cui, X.~Wang, J.~Pei, X.~Yao, and W.~Zhu, ``Arbitrary-order
  proximity preserved network embedding,'' in \emph{Proceedings of the 24th ACM
  SIGKDD International Conference on Knowledge Discovery \& Data Mining}.\hskip
  1em plus 0.5em minus 0.4em\relax ACM, 2018, pp. 2778--2786.

\bibitem{levy2014neural}
O.~Levy and Y.~Goldberg, ``Neural word embedding as implicit matrix
  factorization,'' in \emph{Advances in Neural Information Processing Systems
  (NIPS)}, 2014, pp. 2177--2185.

\bibitem{Fu2019Learning}
G.~Fu, C.~Hou, and X.~Yao, ``Learning topological representation for networks
  via hierarchical sampling,'' in \emph{2019 International Joint Conference on
  Neural Networks (IJCNN)}, July 2019, pp. 1--8.

\bibitem{liao2018attributed}
L.~Liao, X.~He, H.~Zhang, and T.-S. Chua, ``Attributed social network
  embedding,'' \emph{IEEE Transactions on Knowledge and Data Engineering},
  2018.

\end{thebibliography}
%



%

\begin{IEEEbiography}[{\includegraphics[width=1in,height=1.25in,clip,keepaspectratio]{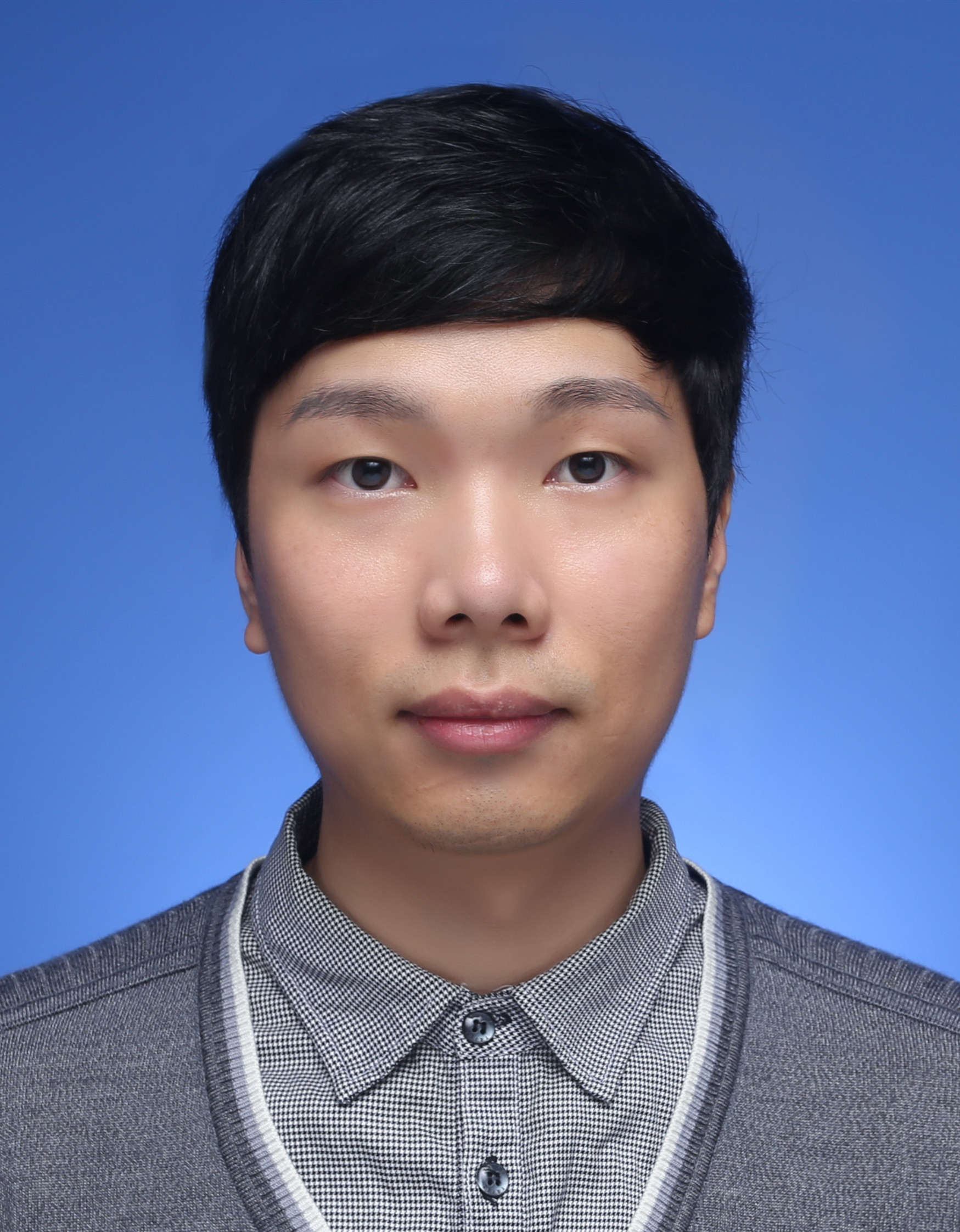}}]{Chengbin Hou} received the B.Eng. first class degree and the M.Sc. distinction degree from University of Liverpool (July 2014) and Imperial College London (November 2015) respectively. He started his Ph.D. in September 2017. Before his PhD study, he worked as an engineer at Huawei. Currently, he is pursuing his Ph.D. degree at Southern University of Science and Technology (SUSTech) and University of Birmingham. He is interested in machine learning and data mining on networked data.
\end{IEEEbiography}

\begin{IEEEbiography}[{\includegraphics[width=1in,height=1.25in,clip,keepaspectratio]{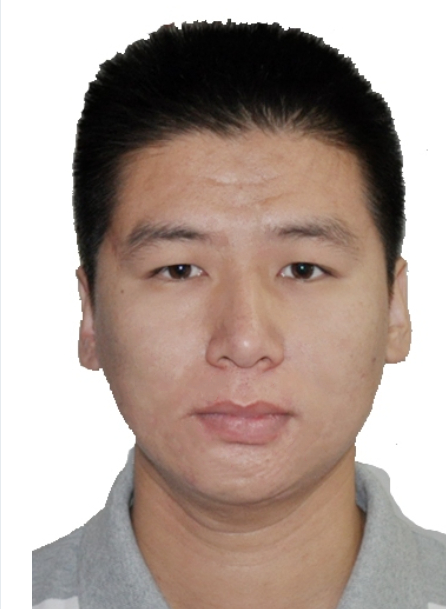}}]{Han Zhang} received his B.Eng. degree in Software Engineering from Xidian University in 2012, then he received MSc degree in Financial Engineering from University of Birmingham in 2014. He is currently pursuing his PhD degree in Computer Science at the University of Birmingham. His current research interests include machine learning and bioinformatics.
\end{IEEEbiography}


\begin{IEEEbiography}[{\includegraphics[width=1in,height=1.25in,clip,keepaspectratio]{Author3}}]{Shan He} received the Ph.D. degree in electrical engineering and electronics from University of Liverpool, Liverpool, U.K., in 2007. He is a Senior Lecturer (Tenured Associate Professor) in School of Computer Science, the University of Birmingham. He is also an affiliate of the Centre for Computational Biology. His research interests include complex networks, machine learning, optimisation and their applications to medicine. He is an Associate Editor of IEEE Transactions on Nanobioscience.
\end{IEEEbiography}

\begin{IEEEbiography}[{\includegraphics[width=1in,height=1.25in,clip,keepaspectratio]{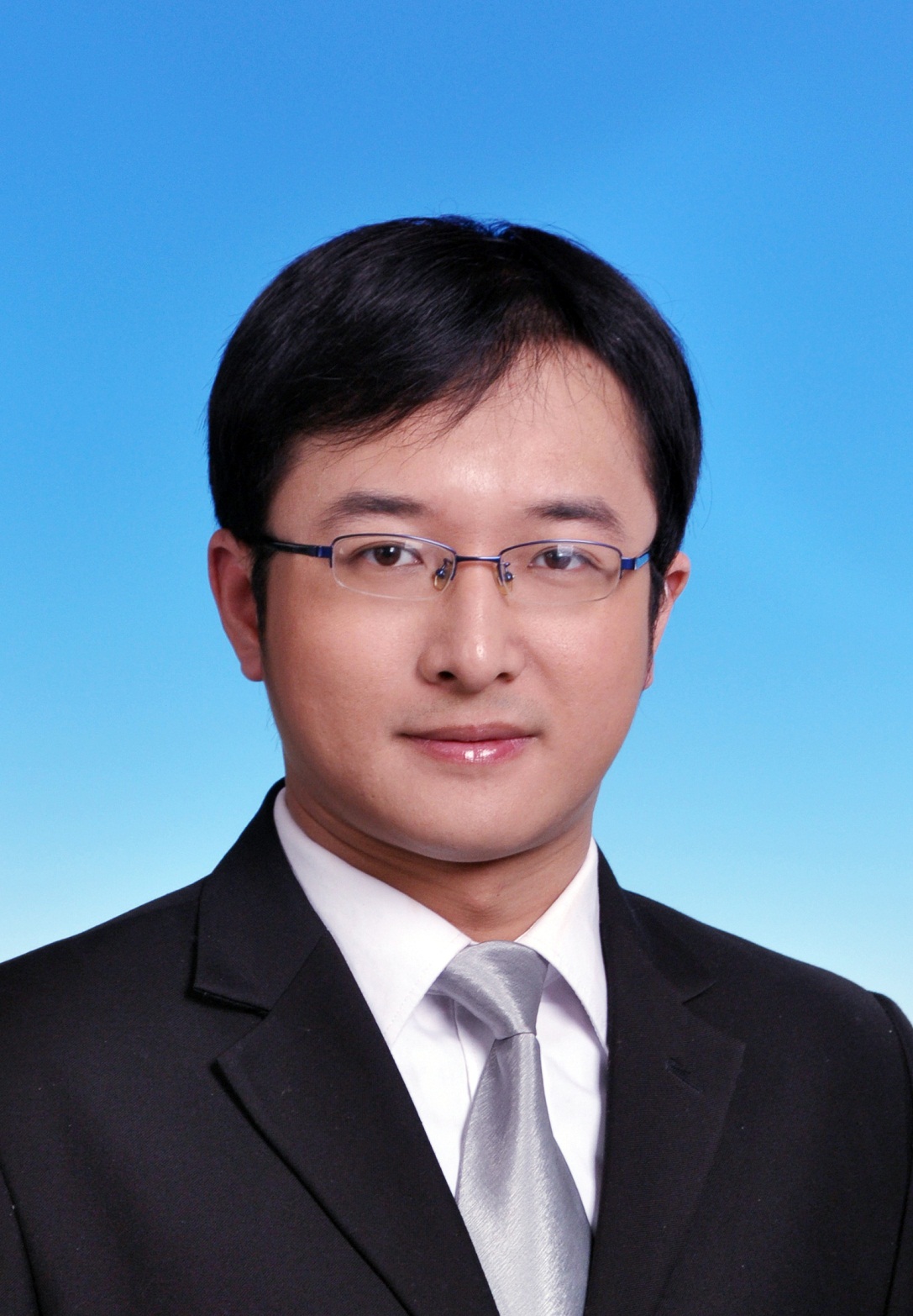}}]{Ke Tang} (Senior Member, IEEE) received the B.Eng. degree from the Huazhong University of Science and Technology, Wuhan, China, in 2002 and the Ph.D. degree from Nanyang Technological University, Singapore, in 2007. From 2007 to 2017, he was with the School of Computer Science and Technology, University of Science and Technology of China, Hefei, China, first as an Associate Professor from 2007 to 2011 and later as a Professor from 2011 to 2017. He is currently a Professor with the Department of Computer Science and Engineering, Southern University of Science and Technology, Shenzhen, China. He has over 9000 Google Scholar citations with an H-index of 45. His major research interests include evolutionary computation, machine learning, and their applications.

Prof. Tang was a recipient of the Royal Society Newton Advanced Fellowship in 2015 and the 2018 IEEE Computational Intelligence Society Outstanding Early Career Award. He is an Associate Editor of the IEEE Transactions on Evolutionary Computation and served as a member of Editorial Boards for a few other journals.
\end{IEEEbiography}




\end{document}